\documentclass[preprint]{aastex}
\bibpunct[ ]{(}{)}{;}{a}{}{,}
\begin{document}

\newcommand{\inv}{$^{-1}$}
\newcommand{\cmsq}{cm$^{-2}$}
\newcommand{\ecsa}{erg cm$^{-2}$ s$^{-1}$ \AA$^{-1}$}
\newcommand{\ecsh}{erg cm$^{-2}$ s$^{-1}$ Hz$^{-1}$}
\newcommand{\ecs}{erg cm$^{-2}$ s$^{-1}$}
\newcommand{\es}{erg s$^{-1}$}
\newcommand{\cxo}{\textit{Chandra\/}}
\newcommand{\hst}{\textit{HST\/}}
\newcommand{\BH}{\bullet}
\newcommand{\LEdd}{L_\mathrm{Edd}}
\newcommand{\Lx}{L_{2-10\,\mathrm{keV}}}
\newcommand{\msun}{M_\sun}
\newcommand{\mbh}{M_\bullet}
\newcommand{\strt}{\rule[-1.5ex]{0pt}{1.5ex}}

\title{Low-Level Nuclear Activity in Nearby Spiral Galaxies}
\author{Himel Ghosh and Smita Mathur}
\affil{Department of Astronomy,The Ohio State University\\140 W 18th Ave, Columbus, OH 43210}
\email{ghosh,smita@astronomy.ohio-state.edu}

\author{Fabrizio Fiore}
\affil{INAF - Osservatorio Astronomico di Roma\\ via Frascati 33, 00040
Monteporzio Catone (Roma), Italy}
\email{fiore@oa-roma.inaf.it}

\and

\author{Laura Ferrarese}
\affil{Herzberg Institute of Astrophysics\\ 5071 West Saanich Road, Victoria, BC V8X 4M6, Canada}
\email{laura.ferrarese@nrc-cnrc.gc.ca}

\begin{abstract}
We are conducting a search for supermassive black holes (SMBHs) with masses
below $\sim\! 10^7\,\msun$ by looking for signs of extremely low-level
nuclear activity in nearby galaxies that are not known to be AGNs. Our survey
has the following characteristics: (a) X-ray selection using the \cxo\ X-ray
Observatory, since x-rays are a ubiquitous feature of AGNs; (b) Emphasis on
late-type spiral and dwarf galaxies, as the galaxies most likely to have
low-mass SMBHs; (c) Use of multiwavelength data to verify the source is an
AGN; and (d) Use of the highest angular resolution available for observations
in x-rays and other bands, to separate nuclear from off-nuclear sources and
to minimize contamination by host galaxy light. Here we show the feasibility
of this technique to find AGNs by applying it to six nearby, face-on spiral
galaxies (NGC 3169, NGC 3184, NGC 4102, NGC 4647, NGC 4713, NGC 5457) for
which data already exist in the \cxo\ archive. All six show nuclear x-ray
sources. The data as they exist at present are ambiguous regarding the nature
of the nuclear x-ray sources in NGC 4713 and NGC 4647. We conclude, in accord
with previous studies, that NGC 3169 and NGC 4102 are almost certainly
AGNs. Most interestingly, a strong argument can be made that NGC 3184 and NGC
5457, both of type Scd, host AGNs.
\end{abstract}

\keywords{galaxies:active---galaxies:nuclei---galaxies:spiral}

\section{Introduction}

The past decade has seen extraordinary improvement in our
understanding of supermassive black holes (SMBHs) --- their growth and
evolution, and their links with their host galaxies. We now realize
that galaxies hosting SMBHs at their centers are the rule rather than
the exception. The observed correlations of the masses $\mbh$ of the
SMBHs with properties of their host galaxies, for example with the
bulge stellar velocity dispersion \citep{fm00,gea00}, show that there
is a close link between the formation and evolution of galaxies and of
the SMBHs they host. Furthermore, a comparison of the SMBH mass
function required to explain the observed luminosity function of
active galactic nuclei (AGNs) with estimates of the local SMBH mass
function \citep[e.g.][]{mea04,sea04} shows that not only must SMBHs be
very common in massive galaxies but that most, if not all, of these
black holes are relics of AGNs active in previous epochs.  Thus
knowledge of the local SMBH mass function enables us to put
constraints on theories of galaxy and SMBH formation and growth, and
AGN lifetimes.

Estimates of the local SMBH mass function are anchored by resolved
stellar or gas dynamical measurements of the masses of $\sim 30$ SMBHs
\citep[see, e.g., the review by][]{ff05}, and otherwise based on the
distribution of host galaxy properties (luminosity of the bulge or
bulge stellar velocity dispersion $\sigma$) and known scaling
relationships between the mass of the SMBH and these properties (most
prominently $\mbh - \sigma$ and $\mbh - M_\mathrm{bulge}$ ; but see
\citealt{gh07b} for a more direct estimate ). Most of the measured
SMBH masses are $\sim 10^8 \msun$ or greater, however, as the sphere
of influence of a less massive SMBH is extremely hard to resolve even
at moderate distances, even with \hst. For example, the sphere of
influence of a $10^6 \msun$ SMBH at 15 Mpc is $\sim 30$
milliarcseconds (mas).  As a result, while different estimates of the
mass function agree for $10^8 \msun \lesssim \mbh \lesssim 10^9 \msun$
, the low-mass end ($\mbh \lesssim 10^6 \msun$) often has
discrepancies \citep[see, e.g., Fig.~7 of ][ for a comparison of
different authors' estimates of the mass function]{gdea07}. A second
source of uncertainty at the low-mass end is the fact that it is
unknown how the scaling relationships extrapolate to very late-type
spiral galaxies, which have little or no bulge component, and to very
low mass galaxies (dE and dSph).  Yet SMBHs \emph{do} exist in very
late-type spirals, e.g.\ NGC 4395, a spiral galaxy of type Sdm, with
$\mbh \sim 3\times 10^5 \msun$ \citep{pea05}, and in very low mass
galaxies, e.g.\ POX 52, a dwarf galaxy, with $\mbh \sim 3\times 10^5
\msun$ \citep{bea04}. Questions that naturally arise at this point
are: Do the scaling relationships break down at low masses? What
determines the mass of an SMBH: the mass of the bulge or the mass of
the dark matter halo? Is there a lower bound to the local SMBH mass
function? A well defined sample of low-mass SMBHs is needed to answer
these questions.

Since we cannot detect low-mass SMBHs by their dynamical signature,
looking for them by signs of their accretion activity may be the only
viable way of detecting them. This of course limits detection to the
subset of low-mass black holes that are active, but this fraction can
be expected to be large, for the following reason. We now understand
that the ``quasar era'' is a function of luminosity, with the space
density of the most luminous quasars peaking at high redshift and that
of lower luminosity quasars peaking at progressively lower redshifts
\citep{fea03,hms05}. This is often referred to as the ``downsizing''
of AGN activity with cosmic epoch.  Moreover, at least some models of
black hole growth \citep{mea04,m04} require anti-hierarchical
growth. That is, higher mass SMBHs attain most of their mass at high
redshift while lower mass SMBHs grow at progressively lower
redshifts. The trends of AGN downsizing and anti-hierarchical growth,
extended logically to the smallest mass SMBHs, imply that these
objects were active in recent times and may still be accreting at the
present epoch.

The Eddington luminosity of a $10^5 \msun$ SMBH is only $\sim \!
10^{43}$ \es; low-mass SMBHs, even if accreting at high rates, will be
low-luminosity AGNs (LLAGNs). The converse is not true; that is, not
all LLAGNs have a low-mass SMBH. In addition to a low SMBH mass, the
low luminosity of an LLAGN may be caused by a low rate or radiatively
inefficient mode of accretion to an SMBH of any mass \citep[present a
study of massive SMBHs accreting at very low
rates]{sfea06,sgea06}. Finally, obscuration may further lower the
observed luminosity. Low-ionization nuclear emission-line region
(LINER) nuclei have been studied at multiple wavelengths
\citep[e.g.][]{escm02,ssd04,dsgs05,fea06} to identify LLAGNs among
them, and these studies have demonstrated that these AGNs are not
necessarily the same type of object with the same physical
characteristics. Since we are interested specifically in the low-mass
end of the SMBH mass function, it is necessary to identify among the
AGNs those that can be expected to have the smallest black holes. One
approach is to use mass estimators based on the luminosity of the AGN
and the width of the broad component of emission lines in its
spectrum. This is the technique that was used by \citet{gh07a} in
constructing their sample of low-mass SMBHs. A second approach, and
the one we use, is to look for AGN activity in galaxies (late-type
spirals, dwarf galaxies) where the known host galaxy-SMBH scaling
relationships predict the lowest-mass SMBHs would reside. This
approach has been used by \citet{svea07,sea08} to find candidate
low-mass SMBHs. However, the scaling relationships are only
statistical and cannot be used to estimate the mass of a particular
SMBH; thus the second approach requires an independent estimate of the
SMBH mass.

A search for active low-mass SMBHs requires confirmation of the
presence of an AGN in each candidate nucleus, and measurement of the
mass of the SMBHs in the confirmed AGNs. The method used by
\citet{gh07a} has the virtue of effectively combining all of the above
into a single step. However, as the luminosity of an AGN decreases,
the optical spectrum of the galaxy nucleus becomes more and more
dominated by host galaxy light, and the signature of the AGN becomes
difficult to detect. Even when the optical spectrum shows no clear
evidence of an AGN, however, such evidence may still be present in
other wavelengths, such as x-ray and radio \citep[e.g.][]{fea04}, and
infrared \citep[e.g.][]{dea06,svea07,sea08} \citep[See][ for a review
of nuclear activity in nearby galaxies]{h08}. For the
lowest-luminosity AGNs, therefore, it is possible that a system based
on optical spectra would not classify the nuclei as AGNs at all. We
choose to use x-ray selection to identify candidate AGNs for the
following reasons: First, x-rays can penetrate obscuring material
which may be hiding the line emitting regions. Second, there are fewer
sources of x-rays in a galaxy than there are of optical and UV
emission and so dilution of the AGN signature by host galaxy light is
less of a problem. Where the luminosity of the AGN is low to begin
with, even a modest amount of absorption may result in the signal
being below the optical background imposed by the galaxy. Third, even
if, as expected in some theories \citep{elb95,n00,nmm03,l03} AGNs that
have luminosities or accretion rates below a cut-off value do not have
broad-line regions, they should still be detectable in x-rays. X-ray
observations have in fact detected AGNs in what were thought to be
``normal'' galaxies \citep[e.g.][]{mea02}. The disadvantage, as
discussed in \S\ref{sec:disc}, is that x-ray observations by
themselves cannot always distinguish between AGNs and other x-ray
sources, such as x-ray binaries (XRBs) and ultraluminous x-ray sources
(ULXs). Multi-wavelength data are needed to determine the type of
source.

As the first step towards assembling a sample of low-mass SMBHs, we
are conducting an x-ray survey of nearby (within 20 Mpc),
\emph{quiescent} spiral galaxies using the \cxo\ X-ray Observatory to
look for nuclear x-ray sources, with an emphasis on late-type
spirals. The high angular resolution of \cxo\ is necessary to ensure
that any detected source is really at the center and is not an
off-nuclear source. The survey is sensitive to an (unobscured) SMBH of
mass $\mbh = 10^4 \msun$ radiating at $\sim\! 2\times 10^{-3}\, \LEdd$
out to the survey limit. We will present details of the survey sample
selection and the \cxo\ observations in a future paper. Here, we
present the methods used and the feasibility of detecting AGNs using
these methods by applying them to six galaxies that meet selection
criteria similar to those used for the survey sample, and for which
x-ray data already exist in the \cxo\ archive.

The paper is organized as follows: \S\ref{sec:sampsel} describes the
criteria used to select the six galaxies presented here; \S\ref{sec:datan}
describes the observations and data analysis, with individual targets
discussed in \S\S\ref{sec:n3169}--\ref{sec:n5457}; the results are
discussed in \S\ref{sec:disc}.

\section{Sample selection\label{sec:sampsel}}

Starting with Third Reference Catalogue of Bright Galaxies\footnote{In
practice, we accessed the RC3 table in the SDSS database.}
\citep[RC3;][]{RC3}, we imposed the following main filters. (1)
Morphological type: we selected spiral galaxies of type Sa through Sd
($ 1.0 \leq T \leq 7.0$); (2) Distance: galaxies that had recession
velocity $cz\leq 3000$ km s$^{-1}$; (3) Galactic latitude: $|b| \geq
30\degr$ to avoid x-ray absorption by gas in our own Galaxy; (4)
Inclination: we required $\log (a/b) \leq 0.4$, where $a$ and $b$ are
the projected lengths of the semi-major and semi-minor axes,
respectively. This was to avoid obscuration by the disk of the host
galaxy; (5) Nuclear inactivity: the galaxy must \emph{not} be known to
be an AGN or starburst. Starbursts were excluded since the vigorous
star formation also increases the likelihood of the presence of XRBs
and ULXs. LINERs were included as there is still debate about whether
the fundamental source of energy in these nuclei is an AGN or star
formation. We also required that the object have the value of the
``goodposition'' flag set to 1 in the RC3 table in the Sloan Digital
Sky Survey (SDSS) database, which indicated that the coordinates of
the galaxy were accurate to 0.1 s in RA and $1\arcsec$ in Dec. All
objects passing these filters were cross-correlated with SDSS Data
Release 3, with a match radius of $30\arcsec$. This resulted in a list
of 76 galaxies. This list was then checked against publicly available
data in the \cxo\ archive. The final sample consists of six galaxies:
NGC 3169 (Sa), NGC 3184 (Scd), NGC 4102 (Sb), NGC 4647 (Sc), NGC 4713
(Sd), and NGC 5457 (Scd).

\section{Data analysis\label{sec:datan}}

Details of the observations that were analyzed are given in
Table~\ref{tab:obs}. X-ray data were downloaded from the \cxo\ archive
and analyzed using version 3.4 of \cxo\ Interactive Analysis of
Observations (CIAO) software. In each case the Level 2 event
list was processed using the observation-specific bad-pixel file and the
latest calibration files, and filtered to exclude times when the
instrument experienced a background flare. The event lists were then
converted to FITS images and the CIAO wavelet source detection tool
\textit{wavdetect} was run on them to determine source positions. For
each observation, source counts were extracted from a circle centered on
the \textit{wavdetect} source position, and with a radius equal to the
greater of 4.67 pixels ($2.3\arcsec$) and the 95\% encircled-energy
radius at 1.5 keV on ACIS-S at the position of the source. Background
counts were taken from an annulus with an inner radius of twice, and an
outer radius of five times, the source circle radius after excising any
sources that happened to fall in the background region.  Counts were
extracted in the 0.3 -- 8.0 keV (Broad), 2.5 -- 8.0 keV (Hard) and 0.3
-- 2.5 keV (Soft) bands. A hardness ratio (HR) was defined as $HR =
(H-S)/(H+S)$, where $H$ and $S$ are the counts in the hard and soft
bands, respectively. Spectral fits were performed with the CIAO tool
\textit{Sherpa}. Uncertainties reported in fit values represent the 90\%
confidence level for one parameter. In all estimations of the AGN Eddington
ratio we have assumed that the bolometric luminosity is ten times the 2--10
keV luminosity of the AGN.

Where available, optical and UV data from HST (WFPC2) were downloaded
from the archive and analyzed using IRAF to extract nuclear fluxes. We
note here that we use ``nuclear flux'' to refer to the total
background-subtracted flux at the location of the nucleus. We use this
flux, and colors if observations in more than one band exist, only as
a consistency check, not to obtain photometry of any putative AGN. At
these extremely low luminosities ($L_\mathrm{bol} \sim
10^{38}$--$10^{40}$ \es), the contribution of the AGN, even if
present, will be a small fraction of the nuclear flux in the optical
and UV bands. In most cases, therefore, we expect the fluxes and
colors to be consistent with the \emph{absence} of an AGN. We check
for anomalous colors or fluxes, since these may indicate the presence
of an additional source of flux, possibly an AGN. As such, we perform
simple aperture photometry for resolved nuclear sources, with the
aperture size matched to the source size.  For point sources
background-subtracted source counts were determined using
\texttt{imexam}, using a 2-pixel radius aperture for the source, and
an annulus with an inner radius of 7 pixels and an outer radius of 12
pixels for the background. These were corrected for CTE effects
\citep{d02} then extrapolated to a $0.5\arcsec$ radius aperture using
the encircled-energy tables of \citet{hea95}.  A correction of $-0.10$
mag was added to represent the magnitude in an infinite aperture.
Proprietary ACS data for NGC 4713 were made available by Martini et
al.\ \citetext{2008, in preparation}. The flux of the resolved nuclear
source was estimated using aperture photometry on the drizzled image.
We use infrared fluxes from the Two-Micron All Sky
Survey \citep[2MASS;][]{2MASS} Point Source Catalog (PSC) for similar
consistency checks.

Individual targets are discussed below. For each target, we first
describe the observed properties in the x-ray, and then in other
wavebands when such observations exist. We then consider whether the
evidence supports the hypothesis that the x-ray source is an AGN.

\subsection{NGC 3169\label{sec:n3169}}

\begin{figure}
\epsscale{0.5}
\plotone{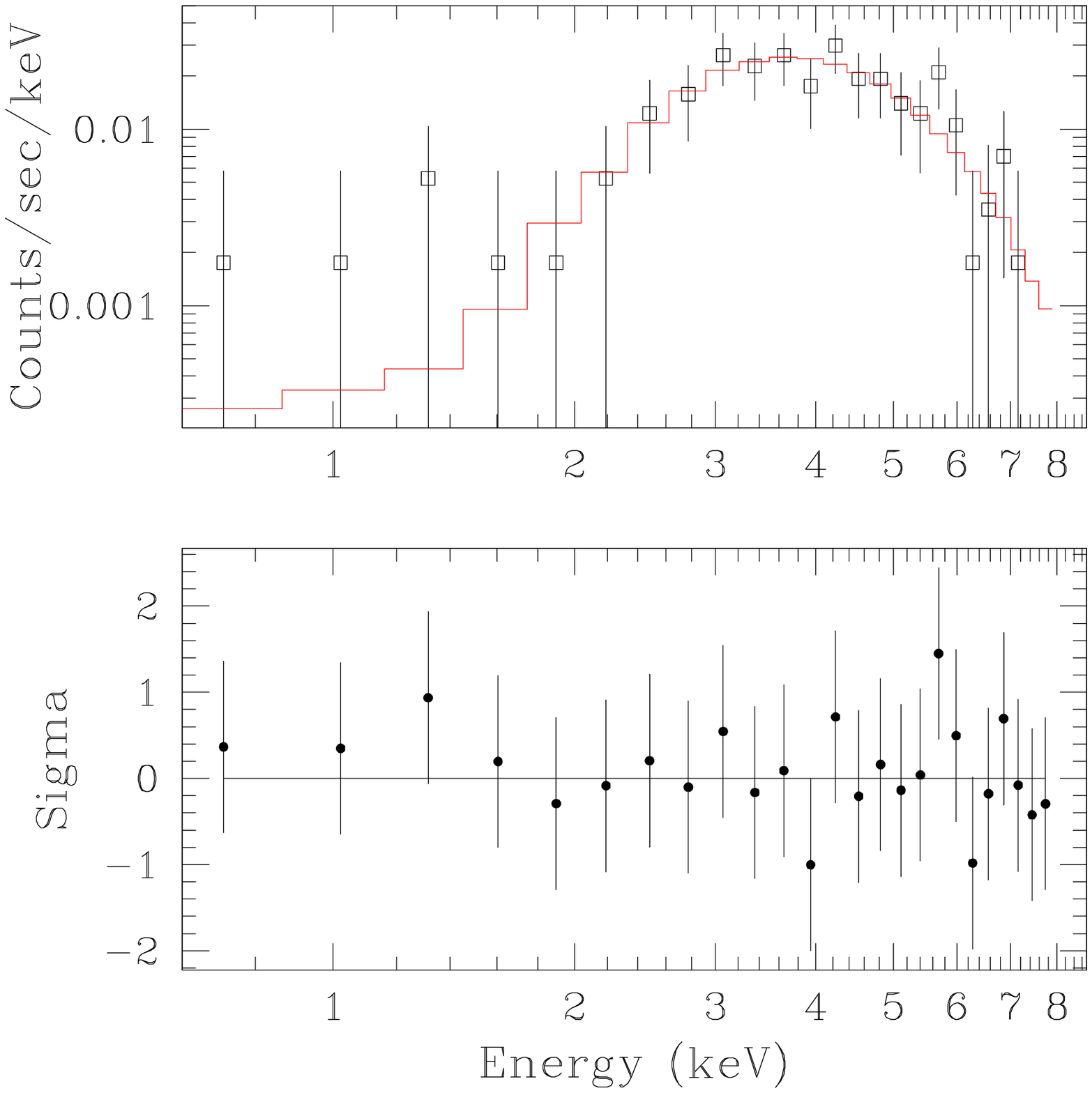}
\caption{Spectrum of NGC 3169. Squares show the data binned by 20 PHA
channels. The solid line shows the best-fit model, an absorbed power-law
with $N_H \sim 10^{23}$ cm$^{-2}$, and $\Gamma = 2.6$. The
lower panel shows the residuals from the fit.}
\label{fig:n3169s}
\end{figure}

This is a galaxy of type Sa, at a distance of 19.7 Mpc \citep{t88},
and its nucleus is classified as a LINER \citep{hfs97-3}. The \cxo\
observation of the nucleus of NGC 3169 was analyzed by \citet{tw03}
and the nucleus identified as a low luminosity AGN, but the x-ray data
are analyzed and presented here again for completeness. NGC 3169 was
observed by \cxo\ for 2 ks and detected with 159 counts. Its hardness
ratio $HR = +0.86$ makes it the hardest of the six sources discussed
here and the only one to have a positive HR. Its spectrum, shown in
Fig.~\ref{fig:n3169s}, can be reasonably fit by an absorbed power-law,
with $N_H \simeq 10^{23}$ \cmsq\ and $\Gamma \simeq 2$. Spectral fit
parameters are shown in Table~\ref{tab:spec}, with the first row
showing a fit using the Cash statistic and data binned to five counts
per bin, and the second row a fit using the Gehrels $\chi^2$ statistic
and data binned by 20 PHA channels. Best-fit parameter values from
both fits are consistent within errors with each other and with the
results obtained by \citet{tw03}. Fluxes used in the analysis are from
the Cash fit. The unabsorbed broad band flux is $f\mathrm{(0.3-8\,
keV)} = 1 \times 10^{-11}$ \ecs, implying a luminosity
$L\mathrm{(0.3-8\, keV)} = 5 \times 10^{41}$ \es\ for the assumed
distance.

Observations with the Very Large Baseline Array (VLBA) show the
nucleus is a mas-scale radio source at 5 GHz \citep{nfw05}. The
nuclear flux density is $f\mathrm{(5\,GHz)} = 6.6 \times 10^{-26}$
\ecsh, or $\nu L_\nu = 1.5 \times 10^{37}$ \es. The measured
brightness temperature exceeds $10^{7.7}$ K and rules out starbursts
and supernova remnants \citep{nfw05}.

The nucleus was detected by 2MASS. The observed magnitudes $J=11.2$,
$H=10.5$, $K_s = 10.0$ imply luminosities $\nu L_\nu \approx (4-6)
\times 10^{42}$ \es\ in these bands.

\subsubsection{The nature of the nuclear emission}

The combination of being a sub-parsec scale radio source as well as a
luminous hard x-ray source points to the source being an AGN. Further
evidence comes from the high luminosity. The source is obscured, as
made clear from the x-ray spectrum. The infrared luminosity therefore
is a better indication of the true bolometric luminosity. This
luminosity, a few $\times 10^{42}$ \es, is higher than is expected
from XRBs or nuclear star formation regions.

\citet{dd06} perform a bulge-disk decomposition of NGC 3169 and use
the 2MASS K-band luminosity of the galaxy to estimate a central black
hole mass $\log\left(\mbh/\msun\right) = 8.2$. \citet{hs98} report the
central stellar velocity dispersion $\sigma_* = 163$ km s$^{-1}$,
which implies $\log\left(M_\BH/M_\sun\right) \approx 7.8$. Assuming
$\log\left(M_\BH/M_\sun\right) \sim 8$ for simplicity, the
corresponding $\LEdd \sim 10^{46}$ \es. The inferred $L_\mathrm{bol}
\sim 10^{42}$ \es\ then implies $L_\mathrm{bol}/L_\mathrm{Edd} \approx
10^{-4}$.

\subsection{NGC 3184\label{sec:n3184}}

\begin{figure}
\epsscale{1.0}
\plottwo{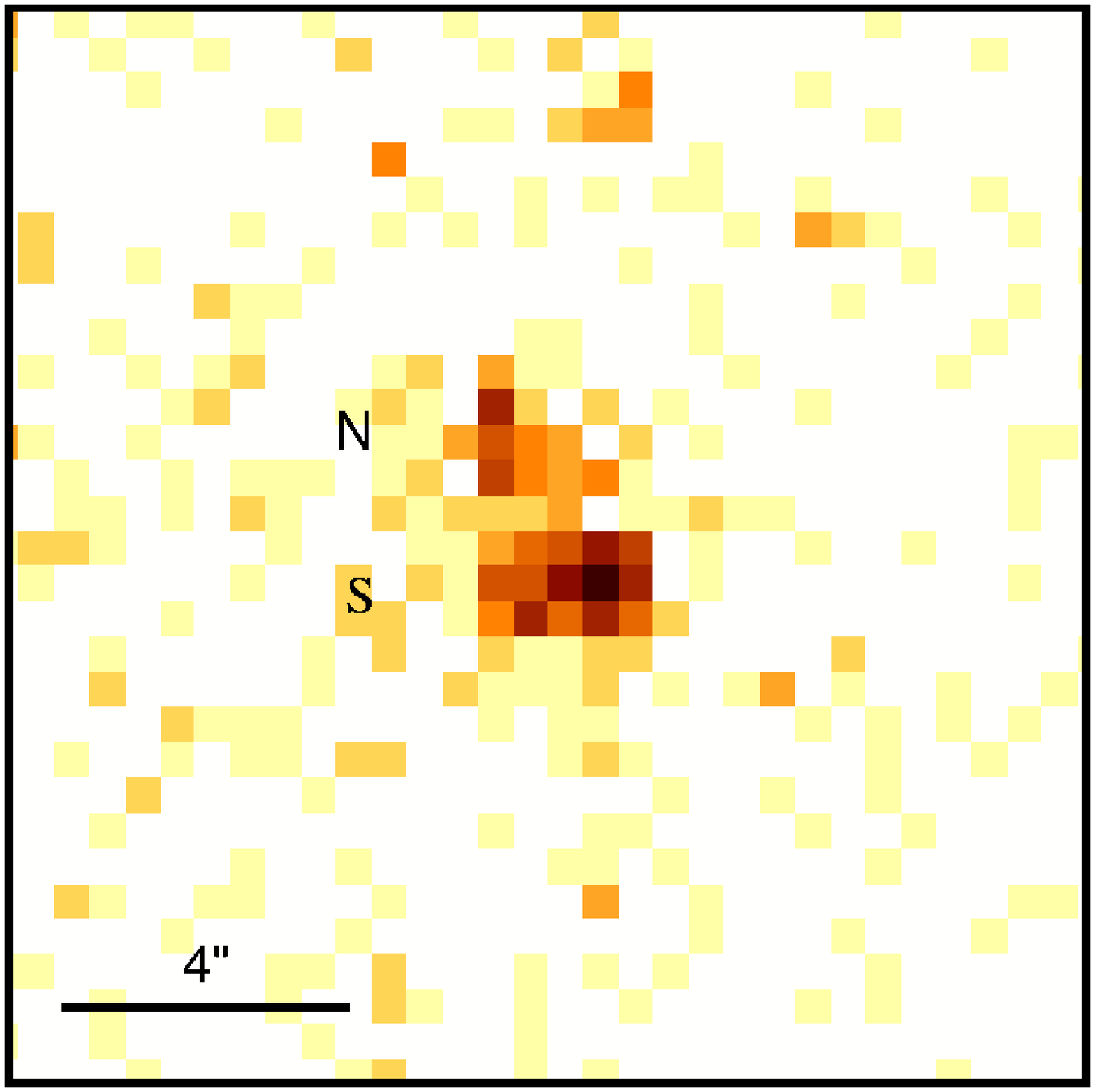}{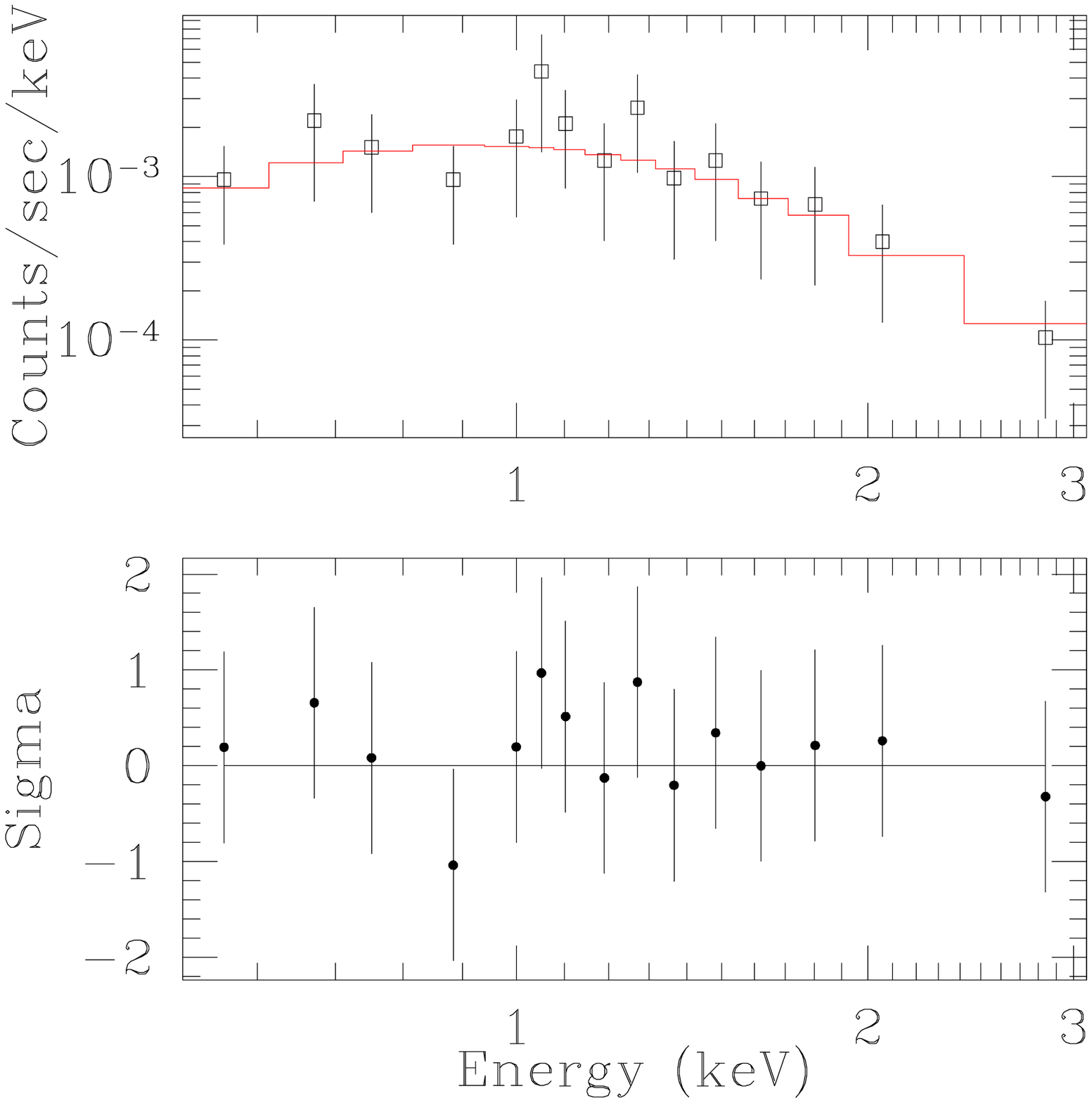}
\caption{\textit{Left:\/} Nucleus of NGC 3184, which appears to
consist of two components, labeled N and S in the image. North is up and
east to the left in the image. About 117 and 36 counts were detected in the
southern and northern components, respectively. The image is $15\arcsec$ on a
side. The black bar in the lower left represents a projected distance of
4\arcsec, corresponding to $\sim\!170$ pc. \textit{Right:\/} Spectrum of the
southern component of the nucleus of NGC 3184, together with an absorbed
power-law model. Data points were binned so that there were at least 5 counts
in each bin. The model shown above (red line) has $N_H = 3.0\times 10^{21}$
cm$^{-2}$ and $\Gamma = 2.6$. The bottom panel shows the residuals.}
\label{fig:n3184i}\label{fig:n3184s}
\end{figure}

This is of type Scd, classified as having an \ion{H}{2} nucleus by
\citet{hfs97-3}, and is at a distance of 8.7 Mpc \citep{t88}. It was
observed with \cxo\ twice, one month apart, for 40 ks and 25 ks. Both
observations used the ACIS-S3 chip. The nuclear x-ray emission appears
to consist of two components (Fig.~\ref{fig:n3184i}), a fainter,
point-like source in the north and a brighter, more extended source in
the south. The total extent of the nuclear emission is about
$4\arcsec$ ($\sim 170$ pc). Both components are soft and have HR $\sim
-0.8$.

Since the northern component has only $\sim 30$ counts, detailed
spectral fitting is not possible. We check for consistency of the data
with different models using unbinned data. An absorbed power-law can
be fitted to the data, and despite the degeneracy in the parameters,
an intrinsic $N_H \gtrsim 6\times 10^{21}$ cm$^{-2}$ is inconsistent
at the 3-$\sigma$ level. Fixing $N_H$ at the Galactic value gives a
range of acceptable power-laws with $1.2 \leq \Gamma \leq 2.8$
(3-$\sigma$ limits). An unabsorbed blackbody with temperature $0.28\,
\mathrm{keV} \leq kT \leq 0.56\, \mathrm{keV}$ (3-$\sigma$ limits) is
also consistent with the data. The observed broadband luminosity of
this component, corrected for Galactic absorption, corresponds to
$L_{0.3-8\,\mathrm{keV}} \sim 2\times 10^{37}$ erg s$^{-1}$. The
difference in count rates between the two observations is not
statistically significant, so this source does not vary on month
timescales.

There are archival \hst\ observations in the UV (WFPC2, F300W) and
optical (WFPC2, F606W) of this galaxy. The F606W image has the nucleus
on the WF4 chip, and clearly shows a central point source. The nucleus
is on the PC chip in the F300W image. There is no central point source
detected. Both images also show diffuse nuclear emission and a
prominent spiral arm in the central $1\arcsec$. Other than the nuclear
source in the F606W image, there are no obvious counterparts of the
x-ray sources in either the UV or the optical. We note here that the
coordinates in the \hst\ images are misaligned with each other and
with the x-ray image. The images were therefore aligned by
cross-correlating point sources in each individual image with SDSS
sources that are classified as stars. The aligned \hst\ images match
each other to within $0.5\arcsec$, and match the \cxo\ image to within
$0.7\arcsec$. These offsets are comparable to the absolute astrometric
accuracy of both \cxo\ and \hst.

\subsubsection{The nature of the nuclear emission}

The fact that we see x-ray emission in the soft band but almost none
in the hard band argues against the emission being the continuum from
the AGN, whether direct or scattered. The source is not seen in the UV
image. Both of these points suggest the presence of heavy
obscuration. The observed x-rays could instead be soft emission from
circumnuclear ionized gas, if the AGN is completely obscured and what
is visible is mostly re-processed radiation. This is in fact the case
in some heavily obscured Seyfert 2s \citep[e.g.][]{bgc06,lea06,gea07},
where the dominant emission is the soft emission from the Narrow-Line
Region. The \cxo\ and \hst\ observations, therefore, are inconclusive
regarding whether the source is an AGN, but do not rule out that
possibility either.

A stronger argument that the source is an AGN derives from infra-red
data. This galaxy is part of the \textit{Spitzer} Infrared Nearby
Galaxies Survey \citep[SINGS;][]{kea03}. \citet{dea06} have used the
equivalent width of the PAH feature at $6.2 \micron$ and the fluxes in
a mix of high- and low-ionization lines ([\ion{S}{4}] $10.51\micron$,
[\ion{Ne}{2}] $12.81 \micron$, [\ion{Ne}{3}] $15.56 \micron$,
[\ion{S}{3}] $18.71 \micron$, [\ion{O}{4}] $25.89 \micron$,
[\ion{S}{3}] $33.48 \micron$, [\ion{S}{2}] $34.82 \micron$) to create
diagnostic diagrams that distinguish between AGN and star-forming
galaxies. This nucleus falls into the ``transition'' region between
AGNs and \ion{H}{2} regions. This suggests an AGN component to the
emission exists that may have been diluted because of the large
aperture used ($\sim 20\arcsec$) to extract the fluxes. In IRAC images
the nuclear source is resolved ($\sim 3.5\arcsec$ FWHM compared to the
$\sim 1.7\arcsec$ FWHM of the PSF). Nuclear fluxes were extracted
using $3\arcsec$ apertures in the IRAC channels (D.\ A.\ Dale, private
communication). Host galaxy emission appears to dominate the MIPS (24,
70, and 160 $\micron$) fluxes, but the observed IRAC colors,
$[3.6]-[4.5] = -0.26$ $\pm 0.16$ and $[5.8]-[8.0] = +0.59$ $\pm 0.16$
(magnitudes in $AB_\nu$ system), are redder than more than 80\% of
normal late-type galaxies \citep{aea08}. This is expected if there is
an AGN, as AGN power-law emission falls off more slowly than galactic
emission in the NIR. Thus, the IR line ratios and IRAC colors strongly
argue for the source to be an AGN.
 
The nucleus was also detected by 2MASS. The $J$, $H$, and $K_s$
magnitudes, from the 2MASS Point Source Catalog (2MASS PSC), are
$J=13.3$, $H=12.7$, and $K_s = 12.5$, which correspond to $\nu L_\nu
\sim 10^{41}$ \es. Such a luminosity can be easily produced by an
AGN. While all the observations are consistent with there being an AGN
in NGC 3184, we cannot rule out a nuclear super-starcluster as the
source. The x-ray emission could be from one or more XRBs in such a
cluster.  The x-ray luminosity ($\sim 10^{37}$ \es) is in the range
seen from XRBs. The presence of a nuclear star formation region is
indicated by the optical line ratios observed by \citet{hfs97-3}. In
addition, \citet{l04} reports a candidate nuclear star cluster in NGC
3184.

In conclusion, there are two scenarios that are consistent with the
observations. In both there is a nuclear star cluster that dominates
the optical emission. In the first, there is no AGN and the x-rays are
produced by one or more XRBs.  In the second, there is a
low-luminosity AGN in the center of the star cluster, so heavily
obscured that we do not observe any direct or scattered emission. The
x-ray emission arises from photoionized gas immediately surrounding
the AGN. Using the luminosity derived using the Bremsstrahlung model
above, and assuming that approximately 1\% of the AGN luminosity is
reprocessed into the plasma x-ray emission, the AGN has a luminosity
$\sim 10^{41}$ \es. The infrared line ratios and IRAC colors from
\textit{Spitzer} strongly argue for this scenario.

\subsection{NGC 4102\label{sec:n4102}}

\begin{figure}
\epsscale{0.5}
\plotone{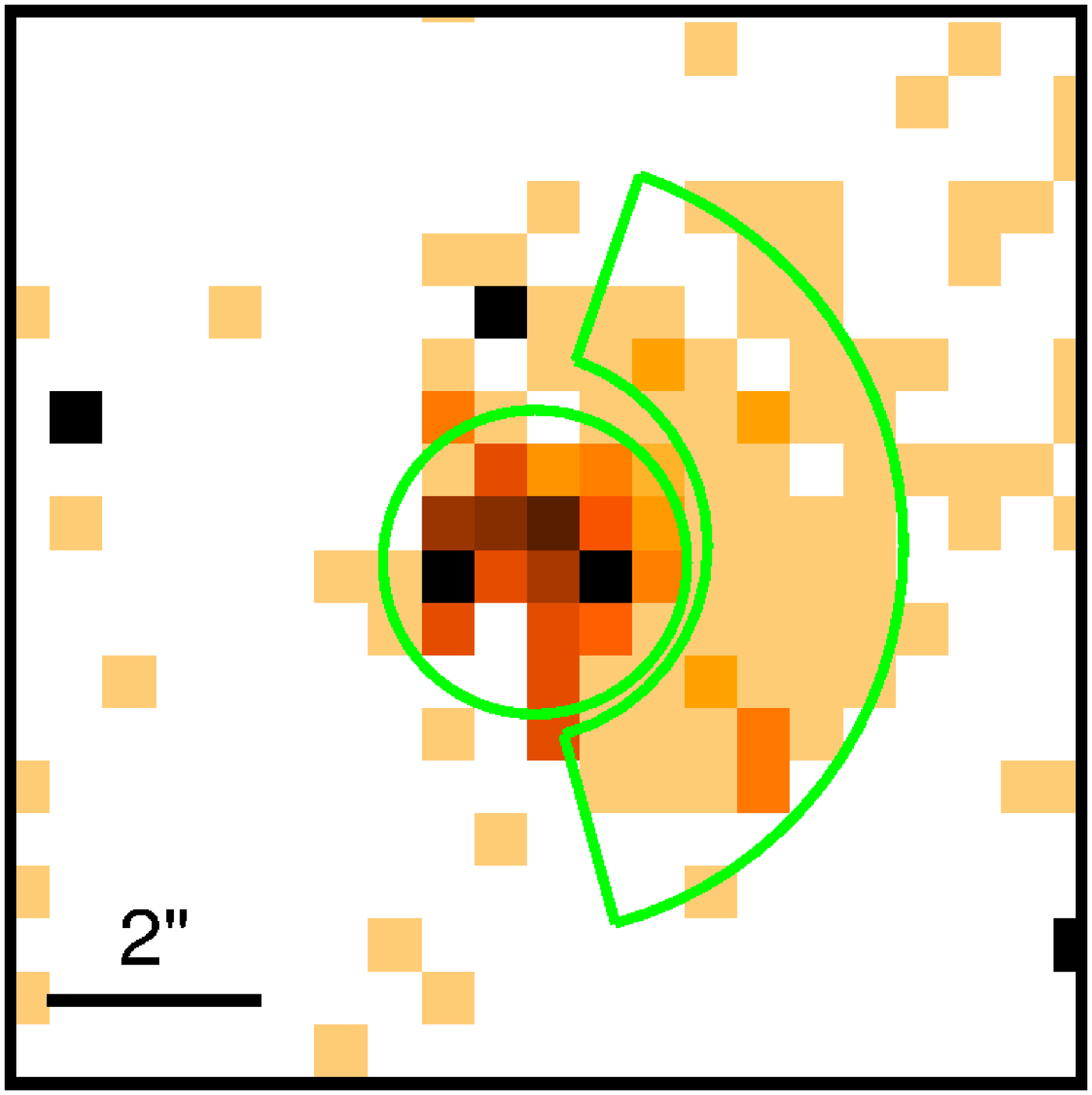}
\caption{Nucleus of NGC 4102 with the regions used to extract counts
overlaid on a hardness ratio map. North is up and east to the left in the
image. The lightest colored pixels have HR $= -1$; the darkest pixels have HR
$= +1$. The circle, 2.9 pixels in radius, was used for the core. The partial
annulus is the region used to extract counts from the extended emission. The
compact core has harder emission than the extended component. The bar on the
lower left represents $2\arcsec$, or a projected distance of $\sim 160$ pc. The
image is $10\arcsec$ on a side.}
\label{fig:n4102reg}
\end{figure}

This is an Sb galaxy, classified as \ion{H}{2} in \citet{hfs97-3} and
as \ion{H}{2}/LINER in the NASA/IPAC Extragalactic Database (NED), at
a distance of 17 Mpc \citep{t88}. The \cxo\ observation of this galaxy
was presented by \citet{dsgs05} and also by \citet{tg07} . There are
about 350 counts in the x-ray image of the nucleus. As a hardness
ratio map (Fig.~\ref{fig:n4102reg}) shows, the hard and soft emission
are well segregated into a fairly hard (HR $=-0.20\pm 0.08$),
point-like source with extended, very soft (HR
$=-0.88^{+0.03}_{-0.06}$) emission to the west of
it. \textit{Wavdetect\/}, run on hard and soft band images, detects a
small hard source and an encompassing soft source whose centers are
$1\arcsec$ apart. We therefore fit the spectra of the harder ``core''
and the softer extended emission separately, using counts from the
regions shown in Fig.~\ref{fig:n4102reg}. The spectrum of the core has
171 counts, and shows a broad line between 6 and 7 keV, suggestive of
reflection. The spectrum was fit after binning to 5 counts per bin,
using the Cash statistic (\texttt{cstat} in \textit{Sherpa}). A simple
absorbed power-law model, fit to the energy range $0.3$--3 keV, shows
increasingly positive residuals at energies above 3 keV, again
suggestive of a reflection component. The fit quality is poor, with a
statistic value of 200 for 27 d-o-f. Using a reflection model
(\texttt{xspexrav}) reduces the statistic to 60 for 26 d-o-f.  The
possible line at 6.4 keV is not well constrained because of the lack
of counts on the high energy side of the line (there are 8 counts with
$E> 7$ keV). For the line, we added a Gaussian at the fixed position
of 6.4 keV and of fixed FWHM $0.3$ keV, allowing only the amplitude to
vary. This reduces the statistic further, to 40 for 25 d-o-f.  The
final parameters are $\Gamma = 2.2^{+0.6}_{-0.5}$ and reflection
scaling factor $R = 129^{+283}_{-85}$.  The formal equivalent width
(EW) of the line is $2.5$ keV, but the uncertainty in the amplitude of
the line is of the order of $\sim 50\%$ and in the normalization of
the reflected component, $\sim 30\%$. The flux, excluding the line and
corrected for Galactic absorption, is $F(0.3\!-\!8\, \mathrm{keV}) =
4.2 \times 10^{-13}$ \ecs. A fit using the $\chi^2$ statistic (shown
in Table~\ref{tab:spec}) gives best-fit parameter values consistent
with those from the Cash fit, with the exception of the equivalent
width of the line, which is $3.3$ keV in this fit. Given the poor
determination of the continuum near the line the difference is not
significant. The equivalent width of the Fe K$\alpha$ in type 1 AGNs
is typically of the order of 300 eV, and a large EW implies that the
direct continuum is suppressed and the source is reflection
dominated. In NGC 4102 the formal EW is an order of magnitude larger;
therefore even after accounting for the large uncertainty we can
conclude that the true EW is greater than the typical type 1 value.

\begin{figure}
\epsscale{1.0}
\plottwo{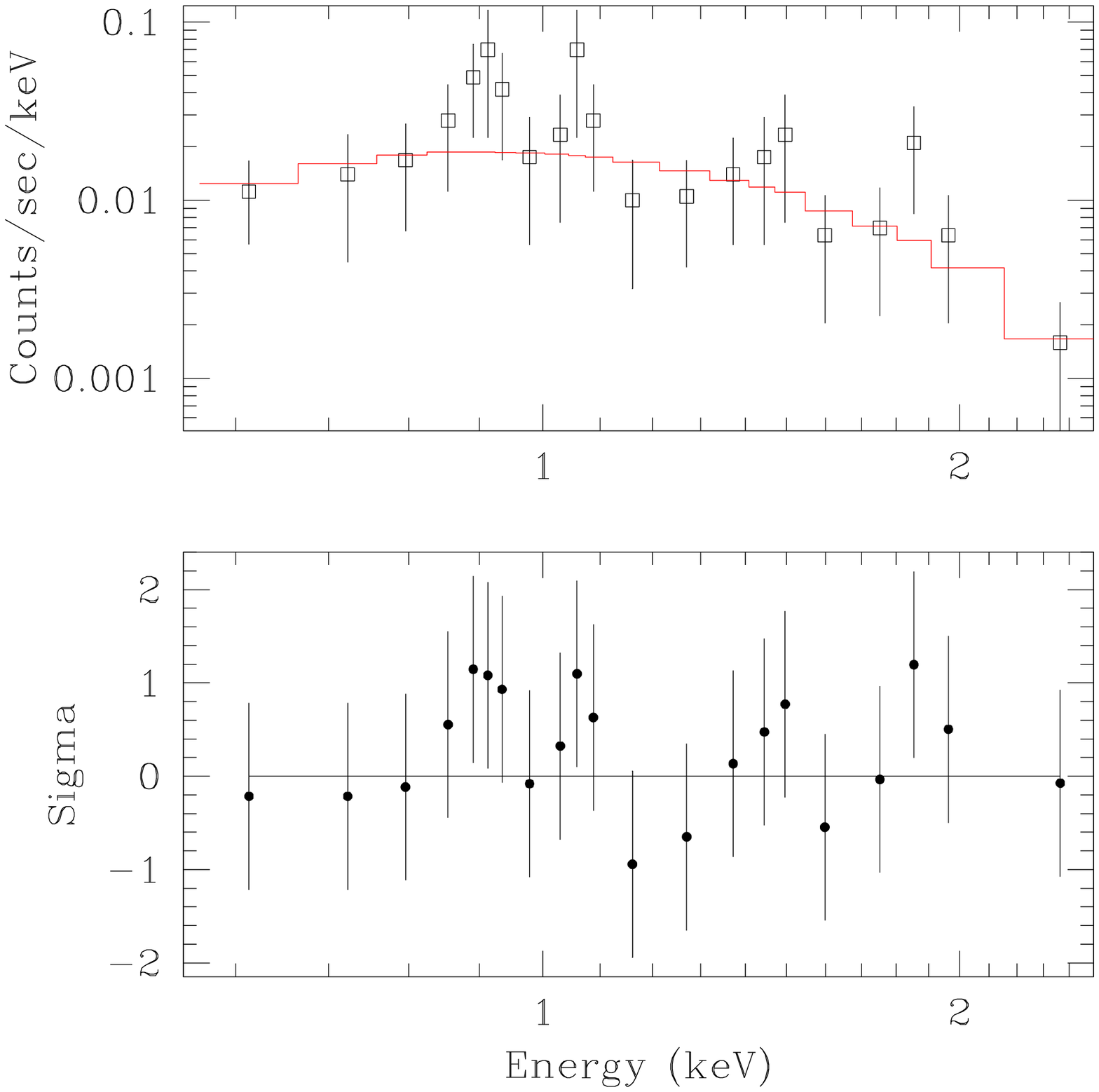}{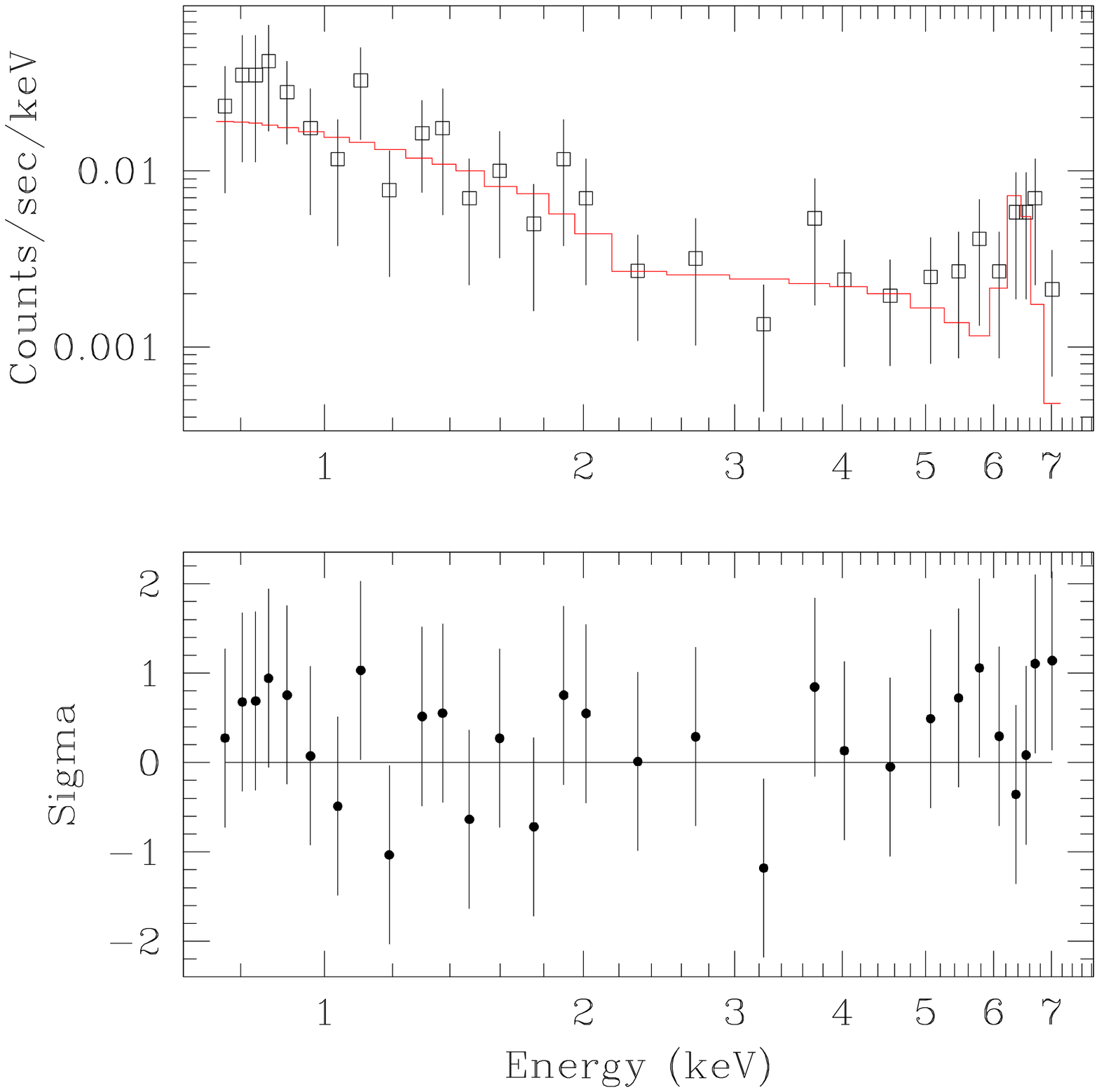}
\caption{Spectra of NGC 4102.  Data have been grouped to 5 counts per bin. \textit{Left:\/} Spectrum of the extended
component. The model is absorbed bremsstrahlung emission, with $N_H = (
1.2^{2.8}_{-1.2}) \times 10^{21}$ cm$^{-2}$ and $kT =
1.2^{+8.6}_{-0.8}$ keV. \textit{Right:\/} Spectrum of the core. The model
is a reflected power-law and a Gaussian line, with no absorption in excess of
Galactic. The best-fit parameters are $\Gamma \approx 2.3\pm 0.6$ and
reflection factor $R\approx 130$. The line was fixed at $6.4$ keV with FWHM =
$0.3$ keV.}
\label{fig:n4102sc}
\end{figure}

The extended emission is very soft, with just 5 counts (out of 115)
above 2.5 keV. The spectrum was fit after binning to 5 counts per bin
and using both the $\chi^2$ and Cash statistics. A bremsstrahlung
model provides an acceptable fit ($\chi^2 = 9.6$ and Cash statistic $=
30$ for 18 d-o-f), and the MEKAL model is less favored ($\chi^2 =
15.6$ and Cash statistic $= 41$ for 18 d-o-f). Best-fit parameter
values obtained using the two statistics are consistent within the
errors. The best-fit temperature $kT \sim 1$ keV is higher than what
may be expected of 100 pc-scale circumnuclear gas. The plasma
surrounding an AGN may be photoionized rather than collisionally
ionized, and there may be complexity in the spectrum that is hidden
because of the poor quality.  The flux in the extended emission is
$F(0.3\!-\!8\, \mathrm{keV}) \approx 2 \times 10^{-13}$ \ecs. Spectral
models and best-fit parameter values for both the core and the
extended emission are shown in Table~\ref{tab:spec}. The $\chi^2$ fits
for both the core and extended components are shown in
Fig.~\ref{fig:n4102sc}.

\hst\ imaging observations in visual and near-infrared (NIR) bands
show that the nucleus is clumpy and has large amounts of dust, and
that there is circumnuclear star formation \citep{csdm97}. High
absorption and reddening towards the nucleus is also indicated by the
Balmer decrement which gives $E(B-V) = 1.00$ \citep{hfs97-3}.  The
nucleus was detected by 2MASS. The 2MASS PSC gives $J = 10.9$, $H =
9.8$, and $K_s = 9.2$. For a distance of 17 Mpc, these correspond to
$\nu L_\nu(J) \approx 6\times 10^{42}$ \es, $\nu L_\nu(H) \approx
8\times 10^{42}$ \es, and $\nu L_\nu(K_s) \approx 7\times 10^{42}$
\es.  The nucleus is highly luminous in the far-infrared
(FIR). Infrared Astronomical Satellite (IRAS) observations indicate
$L_{\mathrm{FIR}} \sim 10^{44}$ \es\ \citep{mt92}, where
$L_{\mathrm{FIR}}$ is the luminosity between $40\micron$ and
$120\micron$, albeit in the arcminute-scale \textit{IRAS} apertures.
The nucleus was also detected in radio by the FIRST survey (1.4
GHz). The stated radio position is within $0.5\arcsec$ of the \cxo\
position, and therefore within \cxo\ astrometric accuracy. The radio
beam size, however, was $3.75\arcsec \times 2.84\arcsec$. The
integrated radio flux density was 223 mJy, with peak flux density of
167 mJy.  This is the same integrated 1.4 GHz flux density reported by
\citet{ccgp82}, fifteen years prior to the 1997 FIRST
observation. \citet{ccgp82} also report a higher resolution 4.9 GHz
observation. The 4.9 GHz map shows radio emission extended in the
northeast-southwest direction with a total extent of about
$5.2\arcsec$ ($\sim\! 420$ pc), and with two peaks separated by
$\sim\! 0.9\arcsec$ ($\sim\! 70$ pc) and also aligned along a NE-SW
axis. The position of the southwest peak coincides with the hard x-ray
``core'' seen in the \cxo\ observation. The radio spectral index
$\alpha = -0.7$ ($S_\nu \propto \nu^\alpha$) suggests a synchrotron
rather than thermal origin for the radio emission.

\subsubsection{The nature of the nuclear emission}

We first consider the case where there is no AGN and all of the
observed emission is due to star formation, and in particular consider
the radio emitting region, which, as mentioned above, has a size of
the order of 400 pc. If this region obeys the radio-FIR correlation
for non-AGN galaxies \citep{c92}, then it has an FIR luminosity of
$\sim\! 10^{44}$ \es, exactly what was measured by
\textit{IRAS}. Therefore the radio and FIR observations are consistent
with each other and with the star-formation-only hypothesis, requiring
only the assumption that this region dominated the FIR emission within
the \textit{IRAS} aperture. Following \citet{c92}, we use the radio
emission to estimate a supernova rate $\nu_{\mathrm{SN}}$ and
consequently a star formation rate (SFR). The luminosity density
measured by FIRST, $L_\nu (\mathrm{1.4 GHz}) \approx 8 \times 10^{28}$
\ecsh, implies $\nu_{\mathrm{SN}} \approx 0.08$ yr\inv, and
$\mbox{SFR}(M \geq 5 \msun) \approx 2 \msun$ yr\inv. In the samples of
\citet{ggs03} and \citet{rcs03}, the galaxies that have this star
formation rate have x-ray luminosities ranging from a few $\times
10^{39}$ \es\ to a few $\times 10^{40}$ \es. The observed nuclear
x-ray luminosity of NGC 4102, $L_X \sim 1.5 \times 10^{40}$ \es\ is
really a lower limit because the amount of absorption is undetermined,
but for the purpose of the argument here may be considered to be
consistent with the $L_X$--SFR relationship of \citeauthor{rcs03} and
\citeauthor{ggs03} On the face of it, therefore, all of the
observations are consistent with a starburst origin to the emission,
but this explanation requires an extraordinarily intense starburst in
a non-interacting galaxy. The radio, FIR, and x-ray luminosities
require $2\, \msun$ yr\inv\ of massive ($M \geq 5 \msun$) star
formation within a 400 pc region. For comparison, the massive star
formation rate in the interacting starburst galaxy M~82 integrated
over the whole galaxy is $2.2\, \msun$ yr\inv\ \citep{nu00}, and the
total SFR ($M \geq 5 \msun$) in the merging pair NGC 4038/4039 is
estimated to be between 5 and 10 $\msun$ yr\inv\ \citep{nu00,ggs03}.

In favor of the AGN hypothesis, NGC 4102 shows an approximately
conical region of outflowing gas that has a higher
[\ion{O}{3}]/H$\beta$ ratio than the surrounding star forming regions
\citep{gfea06}, suggesting exposure to a harder ionizing radiation
than stellar continuum emission, and reminiscent of the ionization
cones sometimes seen in Seyfert 2s \citep[e.g.][]{p88}. The
[\ion{O}{3}] line profile in this region is broader than in the
surrounding regions and cannot be fit with a single Gaussian
\citep{gfea06}. \citet{gvv99} reported a broad component (FWHM
$\approx$ 560 km s\inv) to the [\ion{O}{3}] line as well. The broad
component is weak, comprising just 5--7\% of the flux in the line in
the case of the H$\alpha$ and H$\beta$ lines. Using this component
gives [\ion{N}{2}]$\lambda$ 6583/H$\alpha$ = 1.57, similar to LINERs
and Seyferts, and thus \citeauthor{gvv99} argue for the presence of a
very weak Seyfert 2 in NGC 4102. In this respect NGC 4102 is similar
to NGC 1042, where \citet{swea08} recently demonstrated the existence
of a broad component in [\ion{N}{2}], and that considering only the
broad component moves the nucleus into the Seyfert/LINER regions in
line-ratio diagnostic diagrams.  Finally, the combination of a
point-like hard source, soft extended circumnuclear emission, and an
Fe K$\alpha$ line with a large EW, is one often seen in type 2 AGNs.

There is undeniably strong star formation occurring in the nucleus of
NGC 4102, and we cannot rule out the extremely large SFR implied if
all of the observed emission is imputed to star formation
alone. Nevertheless, the evidence is strong that there is an AGN in
NGC 4102. We conclude that NGC 4102 is another example where an AGN
and strong star formation co-exist at the nucleus.

\subsection{NGC 4647\label{sec:n4647}}

\begin{figure}
\epsscale{0.5}
\plotone{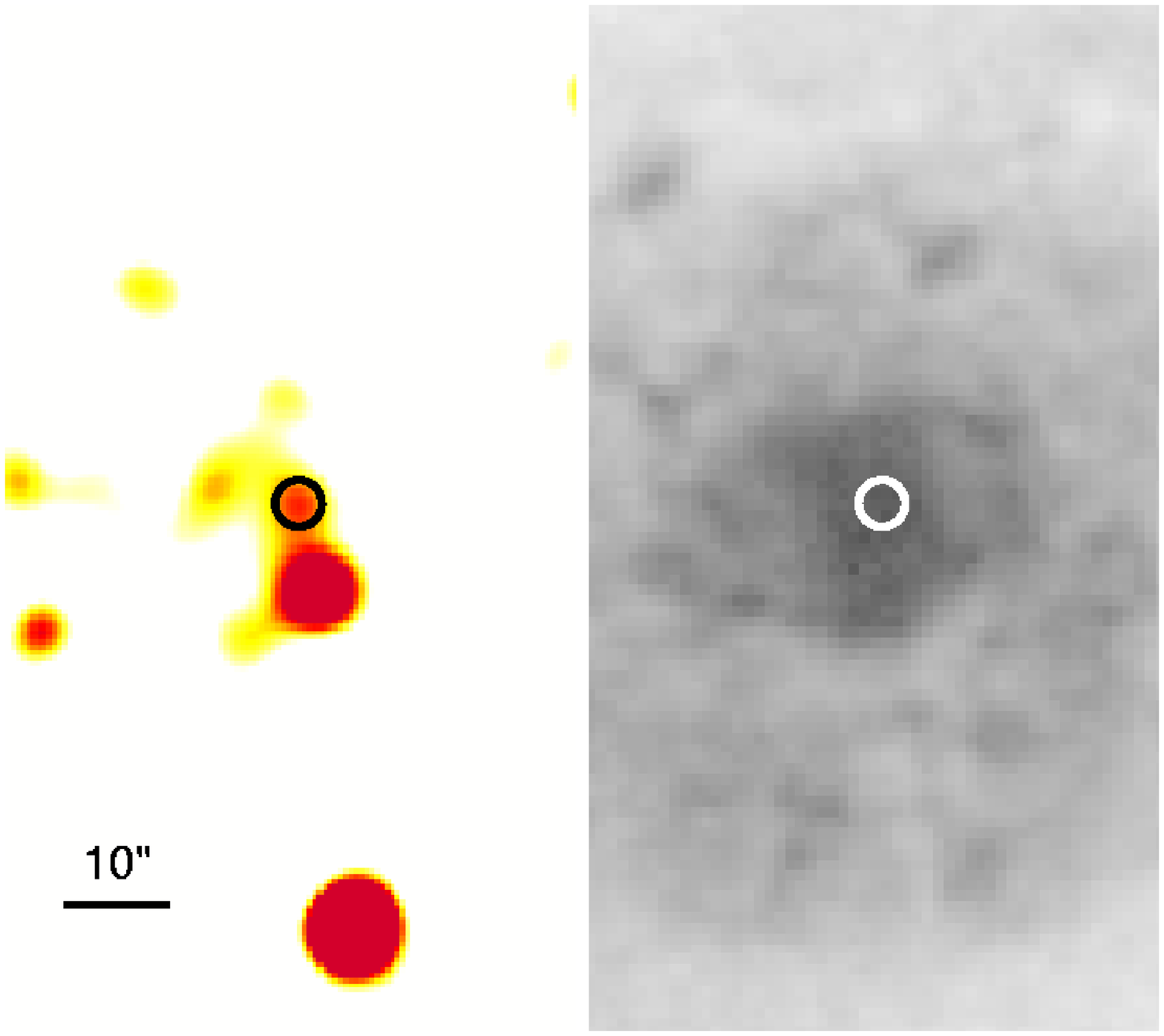}
\caption{On the left are shown the residuals in the \cxo\ image after the
elliptical galaxy NGC 4649 is modeled and subtracted. The residual image
has been smoothed by convolving it with a Gaussian with $\sigma = 5$
pixels. The black circle, whose radius is $2.3\arcsec$ or roughly the
95\% encircled-energy radius at 1.5 keV at that position, shows the
probable nucleus of NGC 4647. On the right, a circle (white) of the same
radius, whose center has the same celestial coordinates as the one on
the left, is superposed on a DSS image of NGC 4647. Both images are shown on
the same scale and have north up and east to the left. The bar in the lower
left corresponds to $10\arcsec$, or a projected distance of $\sim 800$ pc.}
\label{fig:n4647xo}
\end{figure}

NGC 4647 is a galaxy of type Sc, its nucleus classified as \ion{H}{2} in
\citet{hfs97-3}, at a distance of 16.8 Mpc \citep{t88}. The \cxo\ observation
is of the elliptical galaxy NGC 4649, and NGC 4647 is on the chip, at an
off-axis angle of $\sim 2.5\arcmin$. Two factors complicate the detection of
the nucleus of NGC 4647. First, the nucleus lies within the extended emission
from NGC 4649. Second, the nucleus falls on a node boundary of the
CCD. \textit{Wavdetect} at its default filter for source significance
($\sim\!  1$ false source per $10^6$ pixels) does not detect the
nucleus. However, once the elliptical galaxy is modeled and subtracted out of
the image, there is a positive residual at the location of the nucleus of NGC
4647 (see Fig.~\ref{fig:n4647xo}). To extract source counts we used a
circular region centered on the centroid of the residual and with radius
$2.3\arcsec$, which is approximately the 95\% encircled-energy radius at 1.5
keV at that position. To extract background counts we located another region
on the same node boundary that was at the same distance from the center of
NGC 4649 as was the source circle. The source region has $\sim 11$ counts
after background subtraction, but this number necessarily has a large
uncertainty. We take the radius of the source circle, $2.3\arcsec$, as the
\cxo\ positional uncertainty of this source.

The nucleus may also have been detected in x-rays by XMM-Newton
\citep{rsi06}. While the source was detected with $S/N = 11$, the positional
uncertainty was $3\arcsec$. The XMM-Newton source is soft, similar to the
\cxo\ residual. The XMM-Newton and \cxo\ source positions differ by
$5.2\arcsec$. There is a nuclear 2MASS point source ($K_s \approx 12.3$) at a
distance of $1.1\arcsec$ from the \cxo\ position, but it is embedded in
diffuse emission and the flux from the point source is poorly constrained. In
the radio, there is a $5\,\sigma$ upper limit to the nuclear emission of
$0.5$ mJy at 5 GHz
\citep{uh02}. There is an older report of a radio detection of the nucleus
with a flux density of 16 mJy at 1.4 GHz \citep{wod76,k80}, but with
positional uncertainty $(\Delta \alpha, \Delta \delta) = (0.16s, 11.2\arcsec)$.

\subsubsection{The nature of the nuclear emission}

The data are inconclusive at this time as to whether there is an AGN in NGC
4647. The faintness of the source and the positional uncertainties in the
existing observations prevent a firm identification and characterization of
the source.

\subsection{NGC 4713\label{sec:n4713}}

This galaxy is of type Sd, at a distance of 17.9 Mpc \citep{t88}. Its
nucleus was classified as T2 (transition object with type 2 spectrum)
by \citet{hfs97-3}.  The nucleus is clearly detected by \cxo, but with
only ten counts. It is a very soft source, with nine of the ten counts
below $2.5$ keV.  This object was also analyzed by \citet{dsgs05}, but
since they were looking for hard-band sources, this object was not
counted as a detection. The observed count rate corresponds to a flux
of (5--10)$\times 10^{-14}$ \ecs\ in the 0.3--8 keV band, or
luminosity (3--5)$\times 10^{38}$ \es\ for our assumed distance. The
2--10 keV flux depends strongly on the assumed model, from $f_X \sim 1
\times 10^{-17}$ \ecs\ for a thermal bremsstrahlung model with $kT =
0.3$ keV, to $f_X \sim 5 \times 10^{-15}$ \ecs\ for a power-law with
$\Gamma = 2$ and Galactic absorption.

NGC 4713 was observed by \hst\ in December 2006 \citetext{Martini et
al.\ 2008, in preparation}.  The nucleus is resolved; thus only a
nuclear star cluster or star forming region, and no AGN, is
detected. The cluster is $\sim 0.40\arcsec$ in diameter in the image,
corresponding to a physical size of $\sim 35$ pc. Within this aperture
the flux density is $f_\lambda = 5.34 \times 10^{-17}$ \ecsa, which
implies $\nu L_\nu = 1.2 \times 10^{40}$ \es. The nucleus is also in
the 2MASS PSC, with $J$, $H$, $K_s$ luminosities $\nu L_\nu \sim
10^{41}$ \es. However, the nucleus is embedded in diffuse extended
emission and therefore the reported magnitudes may not be accurate
estimates of the nuclear emission. There is an upper limit of $1.1$
mJy to the nuclear radio emission at 15 GHz \citep{nfw05}, and a
similar limit, 1.0 mJy, to the emission at 1.4 GHz from the FIRST
survey.

\subsubsection{The nature of the nuclear emission}

As in the case of NGC 4647, the data are inconclusive regarding the
presence of an AGN. The data are consistent with the AGN hypothesis:
the ``transition object'' classification by \citet{hfs97-3} implies
there may be an AGN component in the optical spectrum; the nucleus is
without doubt an x-ray source, though it is not possible to
distinguish between emission from circumnuclear gas and the nucleus
proper in the current \cxo\ imaging. The hardness ratio (with large
uncertainty) is consistent with a $\Gamma = 2$ power-law. NGC 4713 may
be similar to NGC 3184 in having a low-luminosity AGN inside a nuclear
star cluster.
 

\subsection{NGC 5457 (M 101)\label{sec:n5457}}

\begin{figure}
\epsscale{0.5}
\plotone{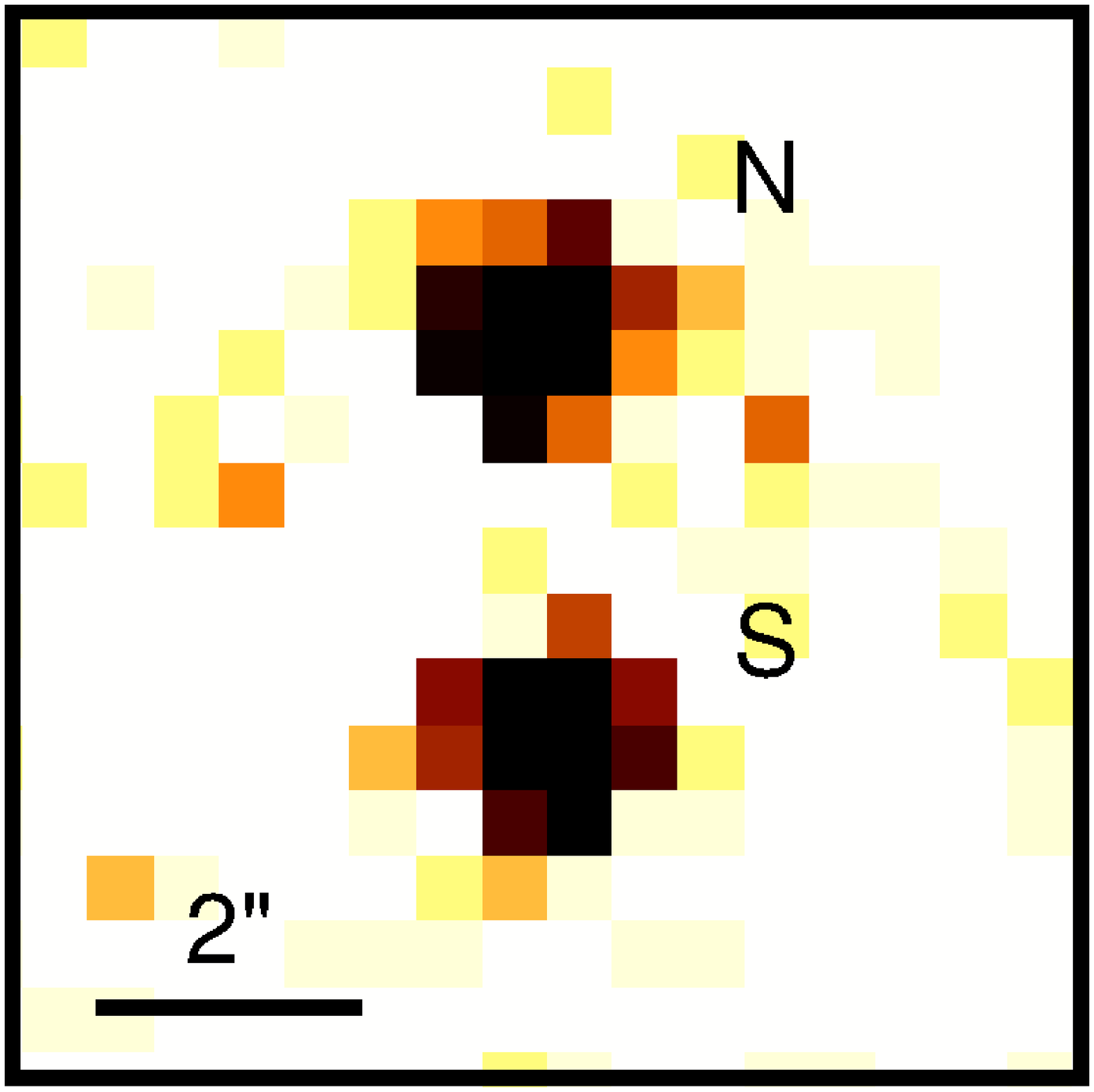}
\caption{Nucleus of NGC 5457 (M 101). North is up and East to the left
in the image. The nucleus is resolved into two sources, marked N and S in the
image. The northern source N is the candidate active nucleus. The southern
source S is a known star cluster \citep{pea01}. The black bar in the lower
left represents a projected distance of 2\arcsec, or $\sim\!70$ pc. The image
is 8\arcsec on a side.}
\label{fig:n5457i}
\end{figure}

\begin{figure}
\includegraphics[angle=-90,scale=0.5]{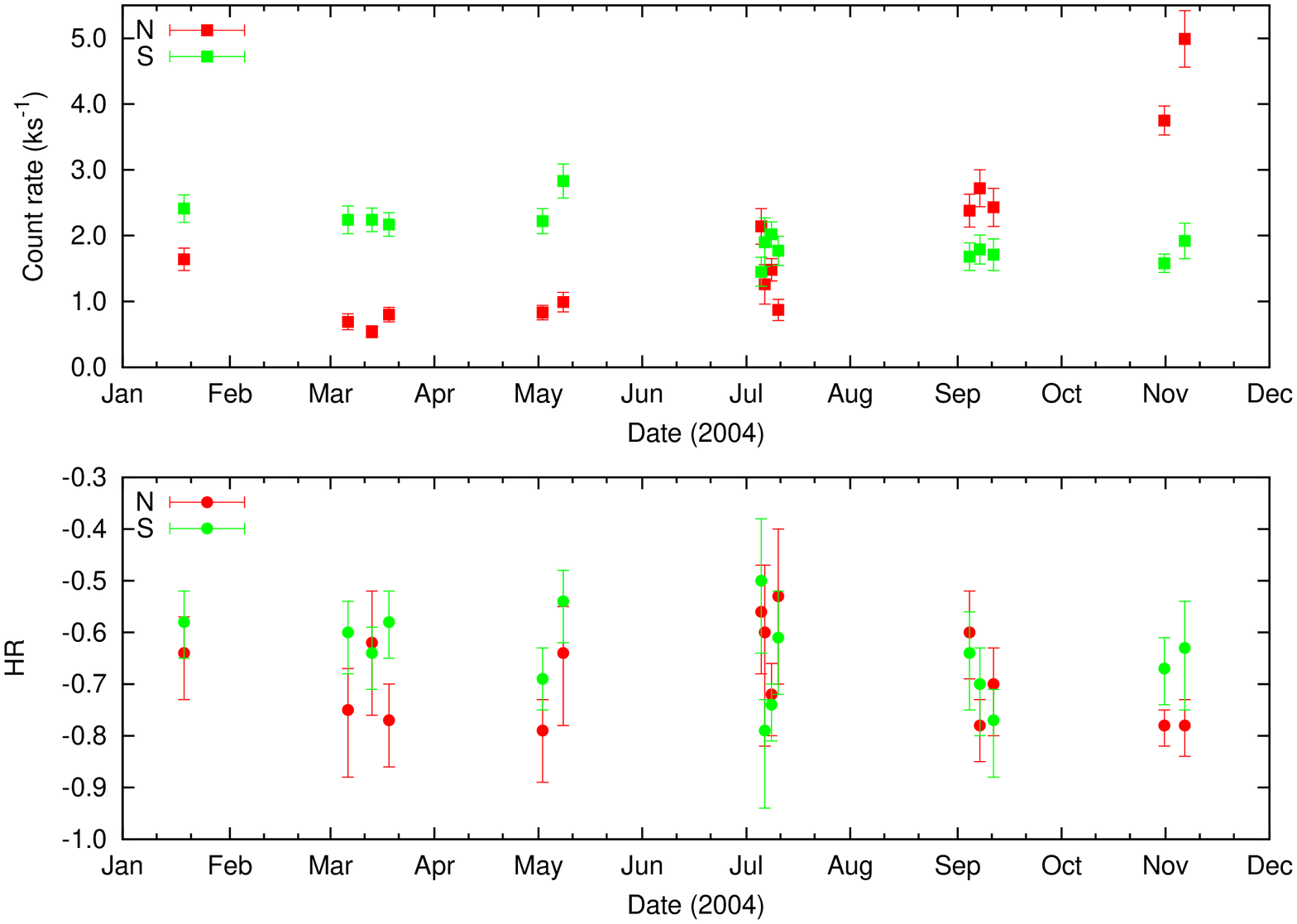}
\caption{Variability and hardness ratio of the two NGC 5457 sources. Only
observations performed in 2004 are shown here. The nucleus (source N) varies
by a factor of $\sim 9$ between March and November. Uncertainty in count rate
was derived assuming $\sqrt{n}$ uncertainty in the counts.}
\label{fig:m101var}
\end{figure}

NGC 5457 is a galaxy of type Scd at a distance of approximately 7 Mpc
\citep{fea01,ssea98}. The nucleus was classified as \ion{H}{2} by
\citet{hfs97-3}. It has been observed multiple times by \cxo\ for a
total observation time of about 1.1 Ms. Our analysis omits the shorter, and
hence low signal-to-noise, exposures. The observations included here are
listed in Table~\ref{tab:m101}. The total usable exposure time is 695
ks. \cxo\ clearly resolves two sources in the nuclear region
(Fig.~\ref{fig:n5457i}), which we label N and S. The northern source, N, is
the nucleus, while the southern, S, is a star cluster
\citep{pea01}. Table~\ref{tab:m101} shows the counts and hardness ratios of
the two sources in each of the observations. Source N (the nucleus) varies in
brightness by about a factor of 9 over the course of about 8 months (see
Table~\ref{tab:m101} and Fig.~\ref{fig:m101var}). There is no significant
change in hardness ratio. In our analysis and discussion below we consider only
the nucleus (source N).

\begin{figure}
\epsscale{0.5}
\plotone{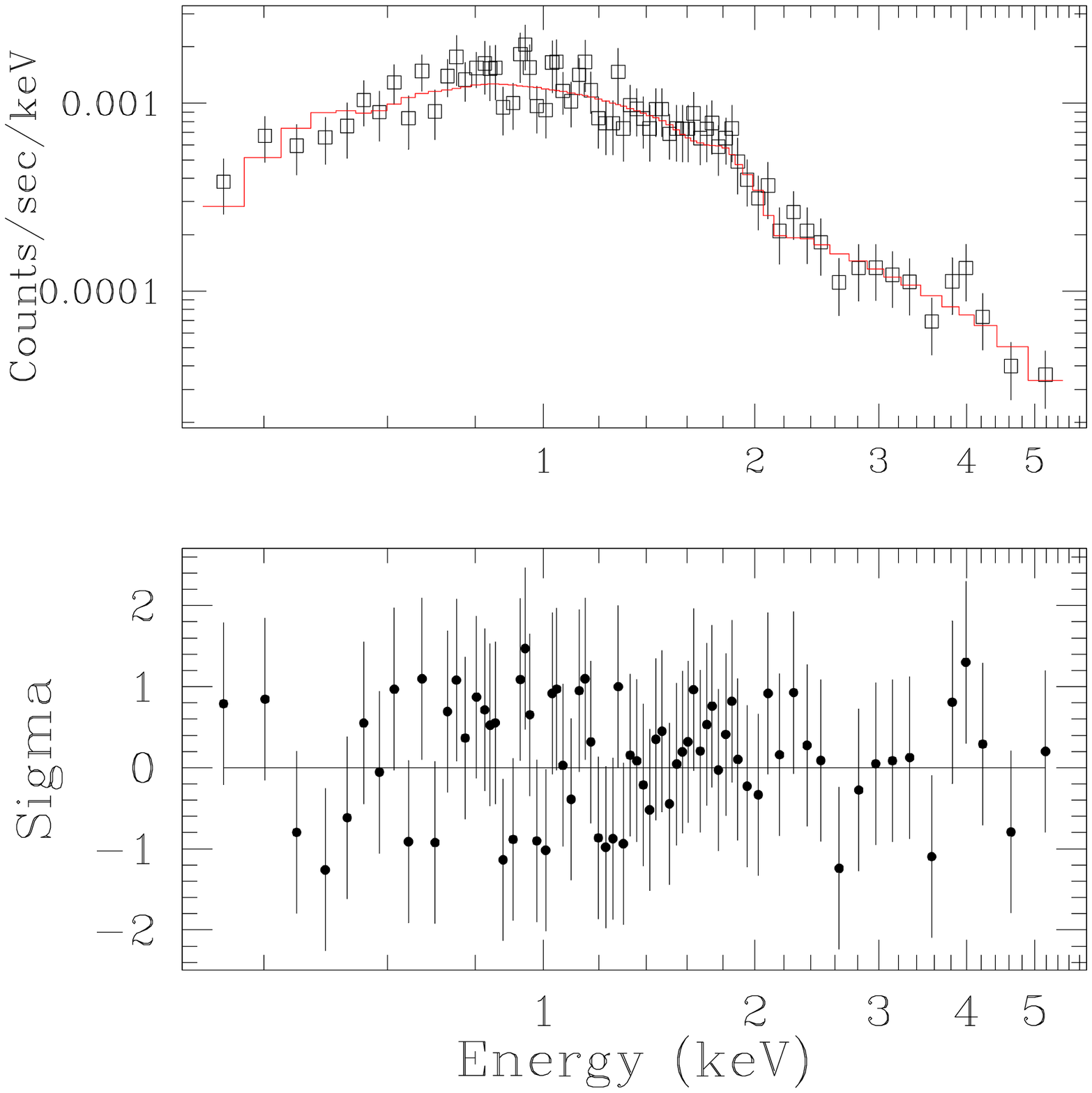}
\caption{Spectrum of NGC 5457 obtained from merging all 2004
observations. Data have been binned to 15 counts per bin. The line shows the
best fit model, an absorbed power-law $(\Gamma = 1.9)$ and thermal plasma (kT =
0.44 keV). 
The line-like feature at 4 keV may be due to statistical fluctuation.
The lower panel shows the residuals.}
\label{fig:n5457sa}
\end{figure}

\begin{figure}
\epsscale{1.0}
\plottwo{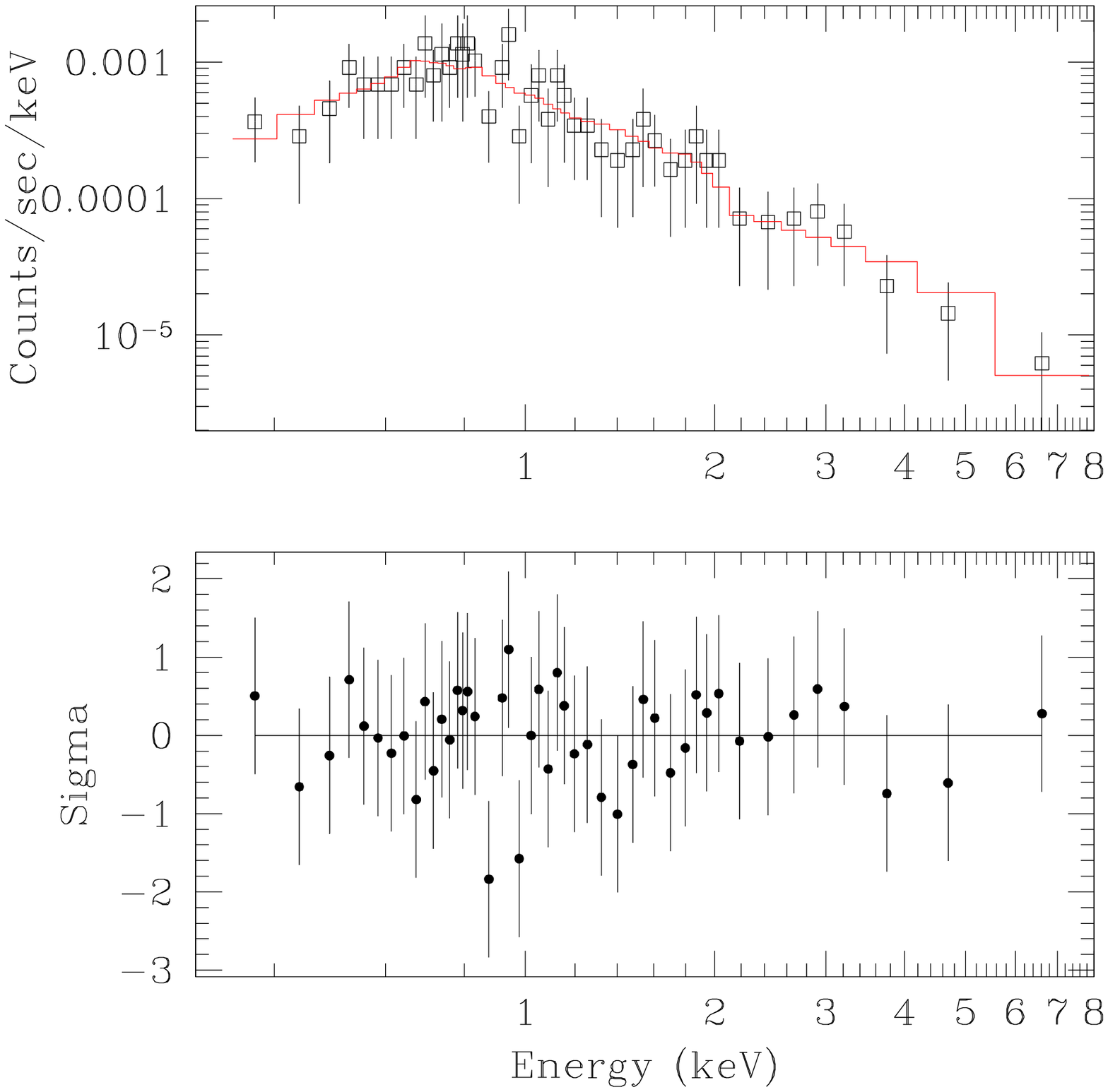}{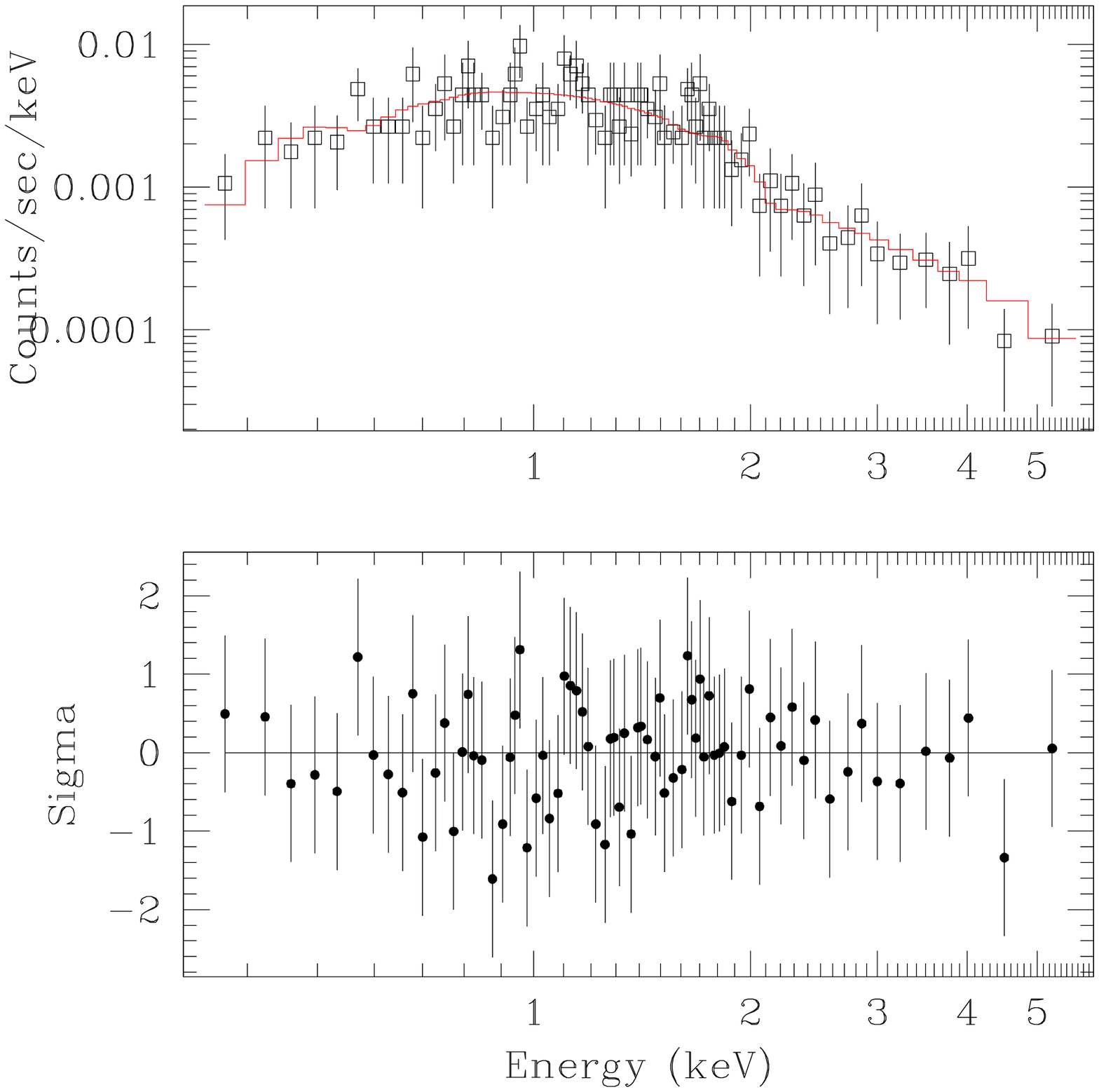}
\caption{Spectrum of NGC 5457 in low and high states. Data have been binned
to five counts per bin. \textit{Left:\/} Low state spectrum.  The best-fit
model is a power-law plus MEKAL plasma with no absorption in excess of
Galactic. \textit{Right:\/} High state spectrum. The best-fit model is a
power-law with both intrinsic and Galactic absorption but no plasma.}
\label{fig:n5457shl}
\end{figure}

\begin{figure}
\epsscale{0.5}
\plotone{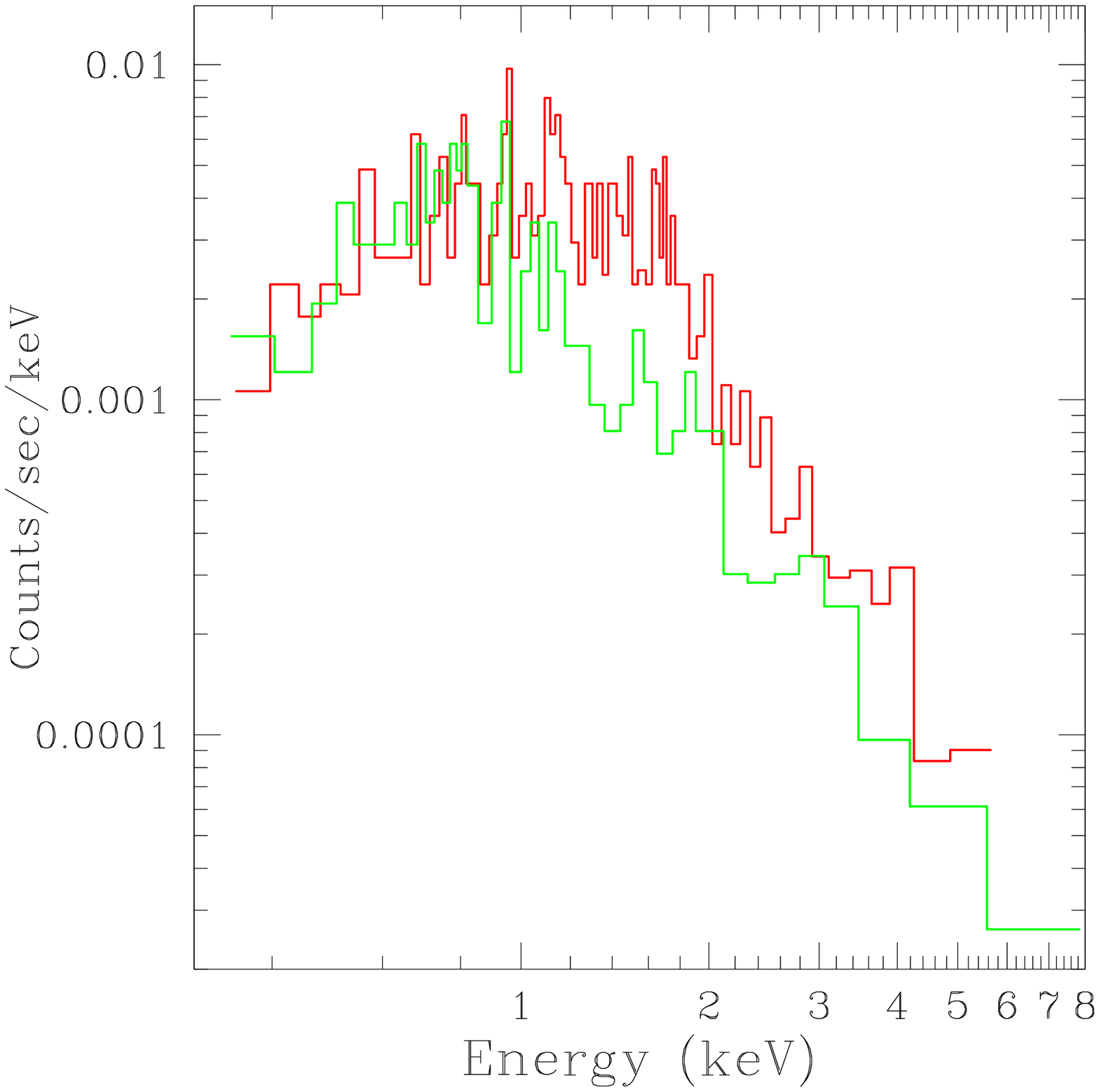}
\caption{Spectrum of NGC 5457 in the high state (red) and the low state
(green) over-plotted.
The spectra are similar below 1 keV. Most of the difference is in the
hard component.
}
\label{fig:n5457sop}
\end{figure}

Three nuclear x-ray spectra were extracted: one (``merged'') from the event
list obtained by merging all observations done in 2004, and fit using the
instrument response from ObsID 5339; one (``high state'') from merging two
observations where the source had a high count rate (ObsIDs 4736 and 6152)
and fit using the response from ObsID 4736; and one (``low state'') from
merging five observations where the count rate was low (ObsIDs 5300, 5309,
4732, 5322, 5323), fit using the response from ObsID 4732. The spectra are
shown in Figs.~\ref{fig:n5457sa} and \ref{fig:n5457shl}; fit models and
parameters are shown in Table~\ref{tab:spec}. The low state spectrum can be
fit with an absorbed power-law ($\chi^2/\mathrm{dof} = 18.7/30$) but the fit
is improved slightly with the addition of a plasma component
(\texttt{xsmekal}) at temperature $kT = 0.3$ keV ($\Delta \chi^2 = -4.5$ for
two fewer d-o-f). The best-fit power-law slope $\Gamma = 1.7\pm 0.5$ is
typical of unabsorbed AGN, and best-fit intrinsic absorption is consistent
with zero.  The unabsorbed 0.3--8 keV luminosity is $\sim 3 \times 10^{37}$
\es. The high state spectrum can also be fit by an absorbed power-law. The
best-fit slope is $\Gamma = 2.2^{+0.4}_{-0.3}$, steeper but consistent with
the low state slope within the uncertainties. The differences from the low
state spectrum are, first, that the fit requires intrinsic absorption $(N_H
\sim 10^{21}$ cm$^{-2}$), and second, that there is no evidence for the MEKAL
component. If a plasma component exists in the high state its flux falls
below that of the power-law component. The unabsorbed 0.3--8 keV luminosity
in the high state is $\sim 3
\times 10^{38}$ \es. Figure~\ref{fig:n5457sop} shows the high and low state
spectra over-plotted. The spectra are identical below 1 keV, with the entire
flux difference arising from the power-law normalization between 1 and 7 keV,
and a possible ``shoulder'' between 1 and 2 keV in the high state. A fit
where the power-law slope is forced to be identical in the low and high
states is of similar statistical significance and produces best-fit values
similar to the separate fits. This fit is shown in Table~\ref{tab:spec} in
the rows labeled ``sim.\ low'' and ``sim.\ high''. The merged spectrum is
similar to the high state one in that an absorbed power-law provides a good
fit without the need for additional components. 

We analyzed archival \hst\ WFPC2 images of NGC 5457, a 2400 s exposure
using the F336W filter (referred to as U band below; nucleus on WF3)
and a 1600 s exposure using the F547M filter (referred to as V band
below; nucleus on PC1). The nucleus and the star cluster are detected
in both images. In 2-pixel radius apertures, the U band nuclear flux
is $6.3 \times 10^{-17}$ \ecsa\ and the V band flux is $8.3\times
10^{-17}$ \ecsa, corresponding to $m_\mathrm{F336W} = 19.3$ and
$m_\mathrm{F547M} = 19.0$ in the STMAG system.  The 2MASS PSC contains
a source at the position of the nucleus (x-ray source N) but none at
the location of source S. The given magnitudes $J=13.1$, $H=12.5$, and
$K_s = 11.8$ all correspond to $\nu L_\nu \approx 10^{41}$ \es.  The
nucleus was not detected by FIRST. \citet{h80} gives an upper limit to
the nuclear luminosity at 6 cm corresponding to $\nu L_\nu < 6 \times
10^{35}$ \es.

\citet{m95} reports $\sigma_* \approx 78$ km s$^{-1}$ which implies
$\log\left(M_\BH/M_\sun\right) \approx 2\times 10^6$. The
corresponding Eddington luminosity is $\LEdd \approx 3\times 10^{44}$
\es. Assuming the source is an AGN, and assuming bolometric correction
factors of $\sim \! 10$ and $\sim \! 1$ in the x-ray and IR,
respectively, implies that the bolometric luminosity is in the range
$10^{39}$--$10^{41}$ \es, or that $L/\LEdd \sim 10^{-5}$--$10^{-3}$.

\subsubsection{The nature of the nuclear emission}

The nuclear source shows several properties typical of an AGN. First,
variability: The source varies by a factor of nine in eight months in
2004, and also varied between 2000 and 2004. Second, the x-ray
spectrum: The spectrum is an absorbed power-law with slope $\sim
2$. In particular, the spectrum cannot be fit by a thermal component
alone --- a power-law is necessary. The low state spectrum is
reminiscent of highly obscured AGN where an unabsorbed but diminished
spectrum is seen via scattering. Third, x-ray colors: The ratio of
0.3--2 keV, 2--5 keV, and 5--8 keV counts puts this object into the
Compton-thick AGN part of the Levenson color-color diagram
\citep[Fig. 9 in][]{lea06}. In the context of the latter two points it
is worth noting that if the source really is a highly absorbed AGN
then energy emitted by the AGN may show up as re-processed radiation
in the infrared. The existence of an infrared source with luminosity
$\sim\!10^{41}$ \es\ is consistent with this picture, though it does not
argue for the presence of an AGN to exclusion of other types of sources.

The x-ray properties of the nucleus are consistent with both the XRB
and AGN hypotheses, but an AGN is not ruled out. The inferred x-ray
luminosity ($10^{37-38}$ \es) is in the range seen from XRBs, and the
observed x-ray colors put this source in the LMXB region of the
\citet{pea03} color-color diagram.  In favor of the XRB scenario is
that the variability of the source is similar to the spectral state
changes in XRBs. The low state may be interpreted as being the XRB
state known as ``hard'' or ``low/hard'', and the best-fit power law
slope, $\Gamma \sim 1.7$ is the value seen in XRBs in this state. The
high state would then be the XRB state known as the ``very high'' or
``steep power law'' state. The best-fit power-law slope $\Gamma \sim
2.2$ in this case is less steep than is usually seen in XRBs in this
state ($\Gamma \gtrsim 2.4$) but the steeper value is included in the
90\% confidence range. However, it must be kept in mind first, that at
the quality of the spectra analyzed here the power law slopes are
consistent with being identical in the two states, and second, that
AGNs can show the same state behavior \citep[e.g.\ 1H
0419-577,][]{pea04}. In XRBs, the steep power-law state is associated
with the presence of quasi-periodic oscillations in the x-ray
emission, while the hard state is associated with the presence of a
radio jet. In principle the presence of these features could provide
additional evidence supporting the XRB hypothesis, but it is not
currently feasible to detect them in XRBs at the distance of NGC 5457.

Thus, there are again two possible scenarios, as in the case of NGC 3184. The
x-ray source could be an HMXB in a super-star cluster. The star cluster would
dominate the optical and IR emission. The large variation observed in the
x-ray flux rules out the source's being more than one HMXB, as otherwise they
would have to be varying in concert.
The actual amount of obscuration is important, however, for the
plausibility of the HMXB hypothesis since the inferred x-ray
luminosity ($\sim 10^{38}$ \es\ in the high state) is already at the
high end of the range of XRB luminosities and there is not much room
for a significantly higher absorption-corrected intrinsic
luminosity. The alternative scenario is an AGN together with a nuclear
star cluster. In AGNs both the intrinsic luminosity and the amount of
obscuration are known to vary. The low state x-ray spectrum may be
explained as a truly under-luminous ($10^{37}$ \es) unabsorbed AGN, or
as an AGN where the obscuration is so high that only scattered light,
and no direct emission, is seen. The fact that the x-ray colors are
similar to those of Compton-thick AGN \citep{lea06} supports the
latter view, but we also note the absence of the $6.4$ keV Fe
K$\alpha$ line in the x-ray spectrum that is often present in the
reflected component.  Though both the HMXB and AGN hypotheses are
possible, on balance the AGN appears to be the more plausible one.


\section{Discussion\label{sec:disc}}

The motivation for this paper was to evaluate the feasibility of
detecting low-mass SMBHs in late-type spiral galaxies, which may still
be accreting at the current epoch and if so should be detectable in
x-rays. The six galaxies studied in this paper are not all late-type,
but span the range Sa--Sd. NGC 3169 and NGC 4102 were regarded as low
luminosity AGNs. None of the remaining four nuclei, however, was known
to have an accreting SMBH, of any mass. NGC 3169 and NGC 4102 are of
type Sa and Sb, which have massive bulges and are expected to have
massive SMBHs. For galaxies of types Scd and Sd, on the other hand,
the lack of a luminous AGN could mean either that there is no SMBH or
that the mass of the SMBH is low. As such, the observations studied
here examine both aspects of a search for accreting low-mass SMBH:
First, are these objects really detectable, given that the accretion
rate is expected to be low? Second, are any sources detected in the
very latest type spiral galaxies that have small or no bulges?

We first note that of the six galaxies presented here, all six show
nuclear x-ray sources.  This implies that it is a very common
occurrence. In a survey of late-type spiral galaxies such as our
ongoing \cxo\ survey, therefore, the predominant concern is not going
to be detection efficiency, but rather identification of the AGNs
among the detected sources. Given that the sample presented in this
paper consists of only six galaxies, we do not draw statistical
conclusions here of the prevalence of very low-luminosity AGNs in
nearby galaxies. But we note that, as shown in \S\S 3.1--3.6, of the
six nuclear x-ray sources, NGC 3169 is almost certainly an AGN, and
NGC 4102, NGC 3184, and NGC 5457 , have very strong, though not
conclusive, arguments in favor of their being AGNs. The two remaining
galaxies, NGC 4713 and NGC 4647, are ambiguous but AGNs are not ruled
out. We discuss below the issues such surveys will face when
attempting to identify the nature of the detected sources. The
diagnostic tools traditionally used to distinguish AGNs from non-AGNs
\citep[e.g.\ optical line ratios,][]{bpt81,vo87} were developed in the
course of studying luminous AGN. Dilution of the AGN emission by host
galaxy light was not a serious problem and observations with low
spatial resolution (several arcseconds) sufficed. In the study of AGN
that are either intrinsically less luminous or are heavily obscured,
however, host galaxy light becomes increasingly problematic, and
surveys relying on optical spectra \citep[e.g.][]{hfs95,gh04} require
careful subtraction of the starlight using galactic spectral
templates. In the weakest AGNs, however, signs of AGN emission may not
be detected by the usual diagnostics.  This problem will persist until
optical observations with angular resolution of 1--10 mas become
possible so that host galaxy light can be effectively excluded, though
it can be mitigated by using regions of the spectrum where host galaxy
emission is negligible, for example very high energy x-rays (tens to
hundreds of keV).  The detection of a compact radio source unresolved
at milliarcsecond-scales, especially if the source is accompanied by
jets, would also unambiguously identify the nucleus as an AGN. This
has been the motivation for radio surveys like that of \citet{nea02}.
Some AGNs obscured in the optical and UV may be detectable using
infrared emission line strengths and ratios
\citep[e.g.][]{dea06,svea07,sea08}.

While the methods listed above allow the unambiguous identification of
AGNs, observations often do not have the angular resolution or
sensitivity to distinguish AGN and non-AGN flux.  Other sources of
radiation in the vicinity of the AGN are, for example, plasma
photoionized by the AGN itself, or a nuclear star cluster. The
targeted AGNs have very low luminosity and thus even moderate amounts
of obscuration may cause a significant decrement in the observed
flux. In most cases, therefore, the AGN contribution should not be
expected to dominate the total observed flux. Consequently,
identifying these AGNs requires a different approach than what can be
used in the case of the more luminous AGNs (Seyferts and QSOs).  AGNs
can be identified using x-ray observations (e.g.\ with \cxo\ and
\textit{XMM-Newton}) solely, but only if they are point sources whose
inferred luminosities are greater than $\sim\! 10^{41}$ \es. Below
that value, AGNs can become indistinguishable from ULXs and XRBs in
x-rays. ULXs can have luminosities of a few times $10^{40}$ \es\
\citep[e.g.][]{sea07} and have x-ray spectra that look similar to AGN
spectra. It has been suggested \citep{srw06} that ULX spectra show a
break at $\sim\! 5$ keV. AGNs are not known to show this break. In
high quality spectra with a large number of counts it may be possible
to exploit this difference to separate ULXs and AGNs.  XRBs have power
law spectra with $\Gamma \sim 2$, can show an Fe K$\alpha$ emission
line, and emit hard x-rays, all characteristics of AGNs as
well. Additionally, even with \cxo's angular resolution, the physical
space probed ranges from the central $\sim\!$ 10 to 100 pc of the
galaxy. The existence of one or more XRBs within that region would be
unsurprising.  However, the fact that the inferred luminosities are as
high as $10^{38}$ \es\ severely constrains the expected number of
XRBs. For example, for NGC 5457 , one of the four galaxies presented
here that are not confirmed AGNs, \citet{pea01} provide a log N-log S
relation as well as the surface density of x-ray point sources as a
function of radius (their Figs.~3 and 4). Approximately 12.5\% of the
sources have luminosities exceeding $10^{37}$ \es. The surface density
of sources in the innermost $0.5\arcmin$ is $\sim 4.75$
arcmin$^{-2}$. Therefore we may expect $\sim\!0.6$ sources
arcmin$^{-2}$ above the luminosity cut-off in the central
$0.5\arcmin$, or $\sim 4\times 10^{-3}$ such sources within the \cxo\
source circle of radius $2.3\arcsec$ that has been used here.  NGC
5457 is of type Scd, and can be taken to be representative of the
other three galaxies, NGC 4647, NGC 3184, and NGC 4713 which are types
Sc, Scd, and Sd, respectively. Thus invoking XRBs and ULXs alone to
explain the x-ray observations leads to physically implausible
conditions, such as requiring that most, or all, quiescent,
non-starburst spiral galaxies have a ULX or $\sim \! 100$ XRBs in the
central $0.5\arcsec-2\arcsec$. Nevertheless, x-ray observations by
themselves will usually be inadequate to distinguish AGNs from
non-AGNs in any particular instance.

Information from other wavebands is thus crucial for the identification
process. For the six galaxies in this paper multi-wavelength photometry is
summarized in Tables~\ref{tab:lum} and \ref{tab:n31spitz}. But here too,
traditional methods of identifying AGNs using flux ratios such as
$\alpha_{OX}$, $\alpha_{KX}$, and $f_X/f_R$, must be used with caution. The
low luminosities of the AGNs mean that the emission in the two bands being
compared may not be from the same object. For instance, for an obscured AGN
surrounded by a nuclear star cluster, the observed x-ray flux may be from the
AGN but the optical flux may be dominated by the cluster.  The existence of
an AGN must instead be inferred by consistency and plausibility checks
considering as much of the spectral energy distribution as possible, and the
goal is the rejection of the hypothesis that all of the observed properties
can be explained without requiring the presence of an AGN. NGC 3184 provides
a good example where different modes of observing, imaging and spectroscopy,
in two wavebands, x-ray and IR, together make a compelling argument for the
presence of an AGN where each observation individually is ambiguous.

Nuclear star clusters are fairly common in spiral galaxies
\citep[e.g.][]{bea02,wea05,salb08}.  A cluster poses two main
problems. First, it makes the presence of XRBs more likely. Second, if
the cluster is young and contains many O and B stars, it may overwhelm
AGN emission in the UV in addition to the optical \citep[e.g. NGC
4303,][]{cgml02}. It may be possible in some cases, as for NGC 1042
\citep{swea08} and NGC 4102 \citep{gvv99}, to attempt a separation of
the cluster and AGN components in the optical spectrum. In addition,
stellar spectra, even for late-type stars, fall faster towards the
infrared than the power-laws typical of AGNs.  The mid-infrared colors
of AGNs, therefore, tend to be redder than those of stellar
populations and this color difference can be used to infer the
presence of an AGN \citep{skea05}.

In addition to the spectral energy distribution, source variability can be a
discriminant, as AGNs are known to vary in all wavelengths, whereas a star
cluster, say, would not. Conversely, if the variation is periodic it would
rule out an AGN and argue for an XRB instead. 

A survey of the type discussed here finds AGNs and strong AGN candidates, but
does not measure the masses of the SMBHs in those AGNs. Measurement of the
mass of one of these SMBHs will be a difficult endeavor, since the sphere of
influence of the black hole cannot be resolved with current technology and
resources. None of the six objects studied in this paper shows broad optical
emission lines whose widths could be used to estimate the BH mass, and this
is likely to be typical behavior. Spectropolarimetry may uncover broad lines
in polarized light in some of the AGNs. Estimates of the SMBH mass may be
made by using known scaling relationships, with the caveat that the
correlations are all based on SMBHs two or more orders of magnitude more
massive than the ones expected to be found by the survey. The least indirect
method is an application of the observed correlation between the x-ray power
law slope and Eddington ratio
\citep{wmp04}. The Eddington ratio and the luminosity in turn provide an
estimate of the mass of the SMBH. Other relationships that can be used are:
(a) the $\mbh$--$\sigma$ relation \citep{fm00,gea00}, but this method becomes
more and more uncertain as the bulge itself becomes ill-defined in the
latest-type spirals; (b) the $\mbh$--$L_\mathrm{bulge}$ relationship
\citep{kr95,md01,mh03}, which has more scatter and also faces the problem of
the definition of the bulge; (c) the $\mbh$--$v_\mathrm{circ}$ relation
\citep{f02,bea03}, which has the advantage that it does not require the
presence of a well-defined bulge; (d) the $\mbh$--$C$ relation \citep{gea01},
where $C$ is the concentration of light. There is also a reported
relationship between black hole mass and core radio power
\citep{sea03,mwea04}, but this relationship is not as well established as the others,
and is based on observations of elliptical galaxies, so its applicability to
the spiral galaxies here is uncertain. Overall, there is unlikely
to be one standard method of measurement that can be applied to these
galaxies, but one or more of the above methods may provide useful estimates
of or limits to the masses of the SMBHs in these AGNs. Mass measurements
independent of the scaling relationships are possible if an object turns out
to have broad emission lines, like NGC 4395, in which case line widths or
reverberation mapping may be used, or if it has maser emission, like NGC
4258, in which case gas dynamics can be used. Mass measurement in other cases
will have to await mas-scale angular resolution in bands other than radio
to resolve the sphere of influence of these black holes.

Of the six galaxies here, NGC 3169, NGC 4102, and NGC 5457 have
measurements of either the stellar velocity dispersion or the
luminosity of the bulge, thus allowing an estimate of their SMBH
masses (assuming here that the source in M 101 is an AGN). The scatter
in the $\mbh$--$\sigma$ and $\mbh$--$L_\mathrm{bulge}$ relations,
together with the uncertainty in the observed flux and the bolometric
correction, result in uncertainties of about an order of magnitude in
the inferred Eddington ratio, but all three objects have $L/\LEdd \sim
10^{-4}$. This is in the range seen in low-luminosity AGNs
\citep[e.g. $L\sim 10^{-5}\LEdd$ for M 81;][]{yea07}, and much higher
than the $L\sim 10^{-9}\LEdd$ seen in truly quiescent SMBHs. These
observations indicate that there is indeed a population of accreting
SMBHs in nearby spiral galaxies that do not show optical signs of
activity but can be uncovered by looking for their x-ray
emission. Such a population will answer the question of whether the
bulge is the dominant component that determines the existence, and
mass, of a nuclear SMBH. The discoveries of AGNs in the Sd galaxies
NGC 4395 \citep{hfsp97} and NGC 3621 \citep{svea07}, and the strong
evidence for AGNs in the Scd galaxies NGC 3184 and NGC 5457 suggest it
is not, at least as far as existence is concerned. Among the SMBHs
discovered in the latest-type spirals (with small or no bulges) and in
the lowest-mass galaxies should be a population of SMBHs with masses
less than $10^6 \msun$, enabling a systematic study of the low-mass
end of the local supermassive black hole mass function.


\begin{deluxetable}{llcccrcccr}
\tablecaption{Targets and observation parameters\label{tab:obs}}
\tablewidth{0pt}
\tabletypesize{\small}
\tablecolumns{10}
\tablehead{
\colhead{Target} & \colhead{Morph.} & \colhead{Nucleus} &
\multicolumn{2}{c}{Coordinates (J2000)} & \colhead{Dist.\tablenotemark{b}} &
\colhead{Scale} & \colhead{Obs. Date} & \colhead{ObsID} & \colhead{Exp.} \\
\colhead{} & \colhead {Type} & \colhead{Type\tablenotemark{a}} & \colhead{RA} & \colhead{Dec} &
\colhead{(Mpc)} & \colhead{(pc/\arcsec)} & \colhead{} & \colhead{} & \colhead{(ks)}
}
\startdata
NGC 3169 & Sa & L2 & 10 14 15.0 & +03 27 58 & 19.7 & 96 &  2001
May 2 & 1614 & 2.0\\
NGC 3184 & Scd & \ion{H}{2} & 10 18 17.0 & +41 25 28 & 8.7 & 42 & 2000 Jan 8 & 804 & 39.8\\
 & & & & & & & 2000 Feb 3 & 1520 & 21.3 \\
NGC 4102 & Sb & \ion{H}{2} & 12 06 23.1 & +52 42 39 & 17.0 & 82 & 2003
Apr 30 & 4014 & 4.9\\
NGC 4647 & Sc & \ion{H}{2} & 12 43 32.3 & +11 34 55 & 16.8 & 81 & 2000 Apr 20 & 785 & 36.9\\
NGC 4713 & Sd & T2 & 12 49 57.9 & +05 18 41 & 17.9 & 87 & 2003 Jan 28 & 4019 & 4.9\\
NGC 5457\tablenotemark{c} & Scd & \ion{H}{2} & 14 03 12.6 & +54 20 57 & 7.2
& 35 & \nodata\tablenotemark{c} & \nodata\tablenotemark{c} & 750\\
\enddata
\tablenotetext{a}{From \citet{hfs97-3}. L2: Type 2
  LINER, T2: Type 2 Transition object.}
\tablenotetext{b}{From \citet{t88} except from \citet{ssea98} for NGC 5457.}
\tablenotetext{c}{NGC 5457 (M 101) was observed multiple times. The total
observation time analyzed here is given in this table. The individual
observations are listed in Table~\ref{tab:m101}}
\end{deluxetable}


\begin{deluxetable}{lrrrrrrcc}
\tablewidth{0pt}
\tablecaption{X-ray Measurements\label{tab:det}}
\tablecolumns{9}
\tablehead{
\colhead{Target} & \multicolumn{6}{c}{Counts} & \colhead{Bkg/Src} &
\colhead{HR\tablenotemark{a}}\\
\colhead{} & \multicolumn{2}{c}{Broad} &
\multicolumn{2}{c}{Hard} &
\multicolumn{2}{c}{Soft} & \colhead{Area Ratio} & \colhead{}\\
\colhead{} & \colhead{Src} & \colhead{Bkg} & \colhead{Src} & \colhead{Bkg} &
\colhead{Src} & \colhead{Bkg} & \colhead{} & \colhead{}}
\startdata
NGC 3169 & 159 & 23 & 148 & 5 & 11 & 18 & 21.4 & \strt $+0.86^{+0.05}_{-0.03}$\\
NGC 3184 N & 36 & 95 & 4 & 33 & 32 & 62 & 92 & \strt $-0.75^{+0.08}_{-0.13}$\\
\phm{NGC 3184} S & 117 & 95 & 13 & 33 & 104 & 62 & 60 & \strt
$-0.77^{+0.05}_{-0.07}$\\
NGC 4102 all & 354 & 48 & 78 & 8 & 276 & 40 & 6.8 & \strt
$-0.55^{+0.04}_{-0.05}$\\
\phm{NGC 4102} core & 171 & 48 & 68 & 8 & 103 & 40 & 52.5 & \strt
$-0.20^{+0.08}_{-0.07}$\\
\phm{NGC 4102} ext & 115 & 48 & 6 & 8 & 109 & 40 & 26.1 & \strt
$-0.88^{+0.03}_{-0.06}$\\
NGC 4647 & 15 & 38 & 1 & 15 & 14 & 23 & 10.1 & \strt
$-0.80^{+0.04}_{-0.20}$\\
NGC 4713 & 10 & 4 & 1 & 0 & 9 & 4 & 21.4 & \strt $-0.69^{+0.09}_{-0.25}$\\
NGC 5457 & 314 & 256 & 23 & 45 & 291 & 211 & 20.8 & $-0.86\pm 0.03$
\enddata
\tablenotetext{a}{Hardness ratio, HR = (H$-$S)/(H+S), where H and S are
the counts in the hard and soft bands respectively, calculated using
the method described in \citet{pea06}. The tool used is available at \url{http://hea-www.harvard.edu/AstroStat/BEHR/}.}
\end{deluxetable}



\begin{deluxetable}{rlcrcc|rcc}
\rotate
\tablewidth{0pt}
\tablecaption{NGC 5457 X-ray Measurements\label{tab:m101}}
\tablecolumns{9}
\tablehead{
\colhead{ObsID} & \colhead{Obs Date} & \colhead{Exp Time} &
\multicolumn{3}{c}{North source} & \multicolumn{3}{c}{South source}\\
\colhead{} & \colhead{} & \colhead{} & \colhead{Counts} & \colhead{Rate} &
\colhead{HR} & \colhead{Counts} & \colhead{Rate} & \colhead{HR}\\
\colhead{} & \colhead{} & \colhead{(ks)} & \colhead{} & \colhead{(ks$^{-1}$)} &
\colhead{} & \colhead{} & \colhead{(ks$^{-1}$)} & \colhead{}}
\startdata
934 & 2000 Mar 26 & $97.8$ & 262 & $2.68$ & $-0.81^{+0.03}_{-0.04}$ & 264 &
$2.70$ & \strt $-0.70 \pm 0.04$\\
4731\tablenotemark{a} & 2004 Jan 19 & $56.2$ & 92 & $1.64$ &
$-0.64^{+0.07}_{-0.09}$ & 136 & $2.41$ & \strt $-0.58^{+0.06}_{-0.07}$\\
5300 & 2004 Mar 7 & $52.1$ & 36 & $0.69$ & $-0.75^{+0.08}_{-0.13}$ & 117 &
$2.24$& \strt $-0.60^{+0.06}_{-0.08}$\\
5309 & 2004 Mar 14 & $70.8$ & 38 & $0.54$ & $-0.62^{+0.10}_{-0.14}$ & 158 &
$2.24$ & \strt $-0.64^{+0.05}_{-0.07}$\\
4732 & 2004 Mar 19 & $69.8$ & 56 & $0.80$ & $-0.77^{+0.07}_{-0.09}$ & 151 &
$2.17$ & \strt $-0.58^{+0.06}_{-0.07}$\\
5322 & 2004 May 3 & $64.7$ & 54 & $0.83$ & $-0.79^{+0.06}_{-0.10}$ & 144 &
$2.22$ & \strt $-0.69 \pm 0.06$\\
5323 & 2004 May 9 & $42.4$ & 42 & $0.99$ & $-0.64^{+0.09}_{-0.14}$ & 120 &
$2.83$ &\strt $-0.54^{+0.06}_{-0.08}$\\
5338\tablenotemark{a} & 2004 Jul 6 & $28.6$ & 61 & $2.14$ & $-0.56^{+0.09}_{-0.12}$ & 41 &
$1.45$ & \strt $-0.50^{+0.12}_{-0.14}$\\
5339\tablenotemark{b} & 2004 Jul 7 & $14.0$ & 18 & $1.26$ & $-0.60^{+0.13}_{-0.22}$ & 27 &
$1.90$ & \strt $-0.79^{+0.06}_{-0.15}$\\
5340 & 2004 Jul 9 & $54.4$ & 81 & $1.48$ & $-0.72^{+0.06}_{-0.08}$ & 110 &
$2.02$ & \strt $-0.74^{+0.04}_{-0.07}$\\
4734 & 2004 Jul 11 & $35.5$ & 31 & $0.87$ & $-0.53^{+0.13}_{-0.17}$ & 63 &
$1.77$ & \strt $-0.61^{+0.09}_{-0.11}$\\
6114 & 2004 Sep 5 & $38.2$ & 91 & $2.38$ & $-0.60^{+0.08}_{-0.09}$ & 64 &
$1.68$ & \strt $-0.64^{+0.08}_{-0.11}$\\
6115 & 2004 Sep 8 & $35.4$ & 96 & $2.72$ & $-0.78^{+0.05}_{-0.07}$ & 63 &
$1.79$ & \strt $-0.70^{+0.07}_{-0.10}$\\
4735 & 2004 Sep 12 & $28.8$ & 70 & $2.43$ & $-0.70^{+0.07}_{-0.10}$ & 49 &
$1.71$ & \strt $-0.77^{+0.06}_{-0.11}$\\
4736\tablenotemark{a}\tablenotemark{b} & 2004 Nov 1 & $77.4$ & 290 & $3.75$ &
$-0.78^{+0.03}_{-0.04}$ & 122 & $1.58$ & \strt $-0.67^{+0.06}_{-0.07}$\\
6152\tablenotemark{a} & 2004 Nov 7 & $26.7$ & 133 & $4.99$ &
$-0.78^{+0.05}_{-0.06}$ & 51 & $1.92$ & \strt $-0.63^{+0.09}_{-0.12}$\\\enddata
\tablenotetext{a}{Source extended, or possibly image smeared.}
\tablenotetext{b}{Possible residual contamination from background flare.}
\end{deluxetable}



\begin{deluxetable}{lccccccccc}
\rotate
\tablewidth{0pt}
\tablecaption{X-Ray Spectral Fits\label{tab:spec}}
\tablecolumns{10}
\tablehead{
\colhead{Target} & \colhead{Galactic} & \colhead{Model\tablenotemark{a}} & 
\multicolumn{6}{c}{Spectral Fit Parameters} &
\colhead{$\chi_\nu^2$(dof)}\\ 
\colhead{} & \colhead{N$_{\mathrm H}$} & 
\colhead{} & \colhead{kT} &
\colhead{N$_{\mathrm H}$} & \colhead{$\Gamma$} & 
\colhead{Refl.\tablenotemark{b}} & \colhead{Line} & \colhead{EW} &
\colhead{}\\
\colhead{} & \colhead{($10^{20}\, \mathrm{cm}^{-2}$)} & \colhead{} &
\colhead{(keV)} & \colhead{($10^{21}\, \mathrm{cm}^{-2}$)} & \colhead{} &
\colhead{} & \colhead{(keV)} & \colhead{(keV)} & \colhead{}
}
\startdata
NGC\,3169 & 2.86 & \texttt{ab(pl)} & \nodata & \strt $99^{+46}_{-36}$ &
$2.0^{+1.2}_{-1.1}$ & \nodata & \nodata & \nodata & $1.1 (26)$\tablenotemark{c}\\
\phm{NGC\,3169} & & \texttt{ab(pl)} & \nodata & \strt $116^{+90}_{-52}$ &
$2.6^{+2.1}_{-1.5}$ & \nodata & \nodata & \nodata & $0.3 (22)$\\


NGC\,4102 core & 1.79 & \texttt{ab(pl)} & \nodata & \strt $0^{+1.4}$  & 
$2.0\pm 0.5$ & \nodata & \nodata & \nodata & $7.4(27)$\tablenotemark{c}\\

 & & \texttt{ab(rn)} & \nodata & \strt $0^{+0.4}$ & $1.8\pm 0.4$ &
 $108^{+149}_{-61}$ & \nodata & \nodata & $2.3 (26)$\tablenotemark{c} \\

 & & \texttt{ab(rn+g)} & \nodata & \strt $0^{+1.9}$ & $2.3^{+0.6}_{-0.5}$ &
 $129$\tablenotemark{d} & $6.4$ & $3.3$ & $0.6 (25)$\\

 & & \texttt{ab(rn+g)} & \nodata & \strt $0^{+0.6}$ & $2.2^{+0.6}_{-0.5}$ &
 $129^{+283}_{-85}$ & $6.4$ & $2.5$ & $1.6 (25)$\tablenotemark{c} \\

\phm{NGC\,4102} ext & & \texttt{ab(br)} & \strt $1.2^{+8.6}_{-0.8}$ & 
$1.2^{+2.8}_{-1.2}$ & \nodata & \nodata & \nodata & \nodata & $0.5(18)$\\

NGC\,5457 sep.\ low & 1.15 & \texttt{ab(me+pl)} & \strt $0.3^{+0.2}_{-0.1}$ &
$0^{+1.8}$ & $1.7\pm 0.5$ & \nodata & \nodata & \nodata &
$0.5(28)$\\ 

\phm{NGC\,5457} sep.\ high & & \texttt{ab(pl)} & \nodata & \strt
$1.5^{+1.1}_{-1.5}$ & $2.2^{+0.4}_{-0.3}$ & \nodata & \nodata & \nodata &
$0.5(40)$\\

\phm{NGC\,5457} sim.\ low & & \texttt{ab(me+pl)} & \strt $0.3^{+0.3}_{-0.1}$
& $0.42^{+1.9}_{-0.42}$ & $2.0^{+0.4}_{-0.2}$ & \nodata &
\nodata & \nodata & $0.5 (28)$\\
\phm{NGC\,5457} sim.\ high & & \texttt{ab(pl)} & \strt \nodata 
& $1.1 \pm 0.6$ & $2.0^{+0.4}_{-0.2}$ & \nodata &
\nodata & \nodata & $0.5 (41)$\\

\phm{NGC\,5457} merged & & \texttt{ab(pl)} & \strt \nodata
& $0.6 \pm 0.4$ & $1.9 \pm 0.2$ & \nodata &
\nodata & \nodata & $0.6 (70)$\\
\enddata

\tablenotetext{a}{Model labels: \texttt{ab}=\texttt{xswabs},
photoelectric absorption; \texttt{ga}=\texttt{xswabs} with value frozen
at Galactic column density towards this target;
\texttt{br}=\texttt{xsbremss}, thermal bremsstrahlung;
\texttt{g}=\texttt{gauss1d}, one-dimensional Gaussian;
\texttt{me}=\texttt{xsmekal}, thermal plasma;
\texttt{pl}=\texttt{powlaw1d}, one-dimensional power law;
\texttt{rn}=\texttt{xspexrav}, power-law reflected by neutral material.}
\tablenotetext{b}{Reflection scaling factor.}
\tablenotetext{c}{Cash statistic (\texttt{cstat} in \textit{Sherpa}).}
\tablenotetext{d}{Unconstrained by fit.}
\end{deluxetable}



\begin{deluxetable}{cccccccc}
\tablewidth{0pt}
\tablecaption{Inferred nuclear luminosities\label{tab:lum}}
\tablecolumns{8}
\tablehead{
\colhead{} & \colhead{Filter} & \colhead{NGC 3169} & \colhead{NGC 4102} &
\colhead{NGC 4647} & \colhead{NGC 3184} & \colhead{NGC 5457} &
\colhead{NGC 4713}\\
\colhead{} & \colhead{or Band} & \colhead{(Sa)} & \colhead{(Sb)} &
\colhead{(Sc)} & \colhead{(Scd)} & \colhead{(Scd)} & \colhead{(Sd)} 
}
\startdata
X-ray & 0.3--8 keV & 41.7 & 40.2 & 39.0\tablenotemark{ab} &
37.3\tablenotemark{c} & 37.5--38.5\tablenotemark{d} & 38.6\tablenotemark{c}\\
UV \& Opt. & F300W & \nodata & \nodata & \nodata & $< 39.4$ & \nodata &
\nodata \\
 & F336W & \nodata & \nodata & \nodata & \nodata & 39.0 & \nodata \\
 & F547M & \nodata & \nodata & \nodata & \nodata & 39.5 & \nodata \\
 & F606W & \nodata & \nodata & \nodata & 39.5 & \nodata & 40.1 \\
IR & $K_s$ & 42.6 & 42.8 & 38.8 & 40.9 & 41.0 & 41.0 \\
Radio & 15 GHz & \nodata & \nodata & \nodata & \nodata & \nodata & $< 36.8$
\\
 & 5 GHz & 37.2 & \nodata & \nodata & \nodata & $< 35.8$ & \nodata \\
 & 1.4 GHz & \nodata & 38.0 & 36.9\tablenotemark{b} & $< 35.1$ & $< 34.9$ &
$< 35.7$ 
\enddata
\tablecomments{Values are $\log(L/\mathrm{erg\:s}^{-1})$ in the
specified bandpass in the x-ray and $\log(\nu L_\nu/\mathrm{erg\:s}^{-1})$
where a single filter, wavelength, or frequency is given. Note that the
values are derived from observations with varied instruments, PSFs,
apertures, and signal-to-noise ratios, and the uncertainties can be as large
as a factor of two.}
\tablenotetext{a}{0.3--12 keV.}
\tablenotetext{b}{{L}arge positional uncertainty makes it unclear whether the
detected source is really the nucleus.}
\tablenotetext{c}{Based on a small number of counts. A power-law spectrum was
assumed. Intrinsic absorption is unknown.}

\tablenotetext{d}{Variable source.}
\end{deluxetable}


\begin{deluxetable}{cccccccc}
\tablewidth{0pt}
\tablecaption{NGC 3184 luminosity in \textit{Spitzer\/}
  bands.\label{tab:n31spitz}}

\tablecolumns{3}
\tablehead{
\colhead{Band} & \colhead{Aperture} & \colhead{$\log(\nu L_\nu)$}\tablenotemark{a} \\
\colhead{$(\micron)$} & \colhead{Radius} & \colhead{(\es)}
}
\startdata
3.6 & $3\arcsec$ & 40.6\\
4.5 & $3\arcsec$ & 40.4\\
5.8 & $3\arcsec$ & 40.7\\
8.0 & $3\arcsec$ & 40.8\\ 
24 & $6\arcsec$ & 41.1\\ 
70 & $10\arcsec$ & 41.5\\ 
160 & $10\arcsec$ & 41.7\\ 
\enddata
\tablenotetext{a}{Uncertainties are 10\% in the $3.6\micron$,
  $4.5\micron$, $5.8\micron$, $8.0\micron$, and $160\micron$ bands;
  4\% at $24\micron$; 12\% at $70\micron$.}
\end{deluxetable}

\acknowledgments

We thank the referee, and P.~Salucci and A.~Seth for helpful comments.
We are grateful to D.~A.~Dale for kindly providing \textit{Spitzer}
fluxes for the nucleus of NGC 3184, and to P.~Martini, T.~B\"{o}ker,
and E.~Schinnerer for providing \hst\ data for NGC 4713 prior to
publication. Support for this work was provided by the National
Aeronautics and Space Administration through Chandra Award Number
GO7-8111X issued by the Chandra X-ray Observatory Center, which is
operated by the Smithsonian Astrophysical Observatory for and on
behalf of the National Aeronautics Space Administration under contract
NAS8-03060. This research has made use of the NASA/IPAC Extragalactic
Database (NED) which is operated by the Jet Propulsion Laboratory,
California Institute of Technology, under contract with the National
Aeronautics and Space Administration. This publication makes use of
data products from the Two Micron All Sky Survey, which is a joint
project of the University of Massachusetts and the Infrared Processing
and Analysis Center/California Institute of Technology, funded by the
National Aeronautics and Space Administration and the National Science
Foundation. 

\clearpage


\begin{thebibliography}{92}
\expandafter\ifx\csname natexlab\endcsname\relax\def\natexlab#1{#1}\fi

\bibitem[{{Assef} {et~al.}(2008){Assef}, {Kochanek}, {Brodwin}, {Brown},
  {Caldwell}, {Cool}, {Eisenhardt}, {Eisenstein}, {Gonzalez}, {Jannuzi},
  {Jones}, {McKenzie}, {Murray}, \& {Stern}}]{aea08}
{Assef}, R.~J. {et~al.} 2008, \apj, 676, 286

\bibitem[{{Baes} {et~al.}(2003){Baes}, {Buyle}, {Hau}, \& {Dejonghe}}]{bea03}
{Baes}, M., {Buyle}, P., {Hau}, G.~K.~T., \& {Dejonghe}, H. 2003, \mnras, 341,
  L44

\bibitem[{{Baldwin} {et~al.}(1981){Baldwin}, {Phillips}, \&
  {Terlevich}}]{bpt81}
{Baldwin}, J.~A., {Phillips}, M.~M., \& {Terlevich}, R. 1981, \pasp, 93, 5

\bibitem[{{Barth} {et~al.}(2004){Barth}, {Ho}, {Rutledge}, \&
  {Sargent}}]{bea04}
{Barth}, A.~J., {Ho}, L.~C., {Rutledge}, R.~E., \& {Sargent}, W.~L.~W. 2004,
  \apj, 607, 90

\bibitem[{{Bianchi} {et~al.}(2006){Bianchi}, {Guainazzi}, \&
  {Chiaberge}}]{bgc06}
{Bianchi}, S., {Guainazzi}, M., \& {Chiaberge}, M. 2006, \aap, 448, 499

\bibitem[{{B{\"o}ker} {et~al.}(2002){B{\"o}ker}, {Laine}, {van der Marel},
  {Sarzi}, {Rix}, {Ho}, \& {Shields}}]{bea02}
{B{\"o}ker}, T., {Laine}, S., {van der Marel}, R.~P., {Sarzi}, M., {Rix},
  H.-W., {Ho}, L.~C., \& {Shields}, J.~C. 2002, \aj, 123, 1389

\bibitem[{{Carollo} {et~al.}(1997){Carollo}, {Stiavelli}, {de Zeeuw}, \&
  {Mack}}]{csdm97}
{Carollo}, C.~M., {Stiavelli}, M., {de Zeeuw}, P.~T., \& {Mack}, J. 1997, \aj,
  114, 2366

\bibitem[{{Colina} {et~al.}(2002){Colina}, {Gonzalez Delgado}, {Mas-Hesse}, \&
  {Leitherer}}]{cgml02}
{Colina}, L., {Gonzalez Delgado}, R., {Mas-Hesse}, J.~M., \& {Leitherer}, C.
  2002, \apj, 579, 545

\bibitem[{{Condon}(1992)}]{c92}
{Condon}, J.~J. 1992, \araa, 30, 575

\bibitem[{{Condon} {et~al.}(1982){Condon}, {Condon}, {Gisler}, \&
  {Puschell}}]{ccgp82}
{Condon}, J.~J., {Condon}, M.~A., {Gisler}, G., \& {Puschell}, J.~J. 1982,
  \apj, 252, 102

\bibitem[{{Dale} {et~al.}(2006){Dale}, {Smith}, {Armus}, {Buckalew}, {Helou},
  {Kennicutt}, {Moustakas}, {Roussel}, {Sheth}, {Bendo}, {Calzetti}, {Draine},
  {Engelbracht}, {Gordon}, {Hollenbach}, {Jarrett}, {Kewley}, {Leitherer},
  {Li}, {Malhotra}, {Murphy}, \& {Walter}}]{dea06}
{Dale}, D.~A. {et~al.} 2006, \apj, 646, 161

\bibitem[{{de Vaucouleurs} {et~al.}(1991){de Vaucouleurs}, {de Vaucouleurs},
  {Corwin}, {Buta}, {Paturel}, \& {Fouque}}]{RC3}
{de Vaucouleurs}, G., {de Vaucouleurs}, A., {Corwin}, Jr., H.~G., {Buta},
  R.~J., {Paturel}, G., \& {Fouque}, P. 1991, {Third Reference Catalogue of
  Bright Galaxies} (Volume 1-3, XII, 2069 pp.~7 figs..~ Springer-Verlag Berlin
  Heidelberg New York)

\bibitem[{{Dolphin}(2002)}]{d02}
{Dolphin}, A.~E. 2002, in The 2002 HST Calibration Workshop : Hubble after the
  Installation of the ACS and the NICMOS Cooling System, Proceedings of a
  Workshop held at the Space Telescope Science Institute, Baltimore, Maryland,
  October 17 and 18, 2002. Edited by Santiago Arribas, Anton Koekemoer, and
  Brad Whitmore. Baltimore, MD: Space Telescope Science Institute, 2002.,
  p.301, ed. S.~{Arribas}, A.~{Koekemoer}, \& B.~{Whitmore}, 301

\bibitem[{{Dong} \& {De Robertis}(2006)}]{dd06}
{Dong}, X.~Y., \& {De Robertis}, M.~M. 2006, \aj, 131, 1236

\bibitem[{{Dudik} {et~al.}(2005){Dudik}, {Satyapal}, {Gliozzi}, \&
  {Sambruna}}]{dsgs05}
{Dudik}, R.~P., {Satyapal}, S., {Gliozzi}, M., \& {Sambruna}, R.~M. 2005, \apj,
  620, 113

\bibitem[{{Eracleous} {et~al.}(1995){Eracleous}, {Livio}, \& {Binette}}]{elb95}
{Eracleous}, M., {Livio}, M., \& {Binette}, L. 1995, \apjl, 445, L1

\bibitem[{{Eracleous} {et~al.}(2002){Eracleous}, {Shields}, {Chartas}, \&
  {Moran}}]{escm02}
{Eracleous}, M., {Shields}, J.~C., {Chartas}, G., \& {Moran}, E.~C. 2002, \apj,
  565, 108

\bibitem[{{Ferrarese}(2002)}]{f02}
{Ferrarese}, L. 2002, \apj, 578, 90

\bibitem[{{Ferrarese} \& {Ford}(2005)}]{ff05}
{Ferrarese}, L., \& {Ford}, H. 2005, Space Science Reviews, 116, 523

\bibitem[{{Ferrarese} \& {Merritt}(2000)}]{fm00}
{Ferrarese}, L., \& {Merritt}, D. 2000, \apjl, 539, L9

\bibitem[{{Filho} {et~al.}(2004){Filho}, {Fraternali}, {Markoff}, {Nagar},
  {Barthel}, {Ho}, \& {Yuan}}]{fea04}
{Filho}, M.~E., {Fraternali}, F., {Markoff}, S., {Nagar}, N.~M., {Barthel},
  P.~D., {Ho}, L.~C., \& {Yuan}, F. 2004, \aap, 418, 429

\bibitem[{{Fiore} {et~al.}(2003){Fiore}, {Brusa}, {Cocchia}, {Baldi},
  {Carangelo}, {Ciliegi}, {Comastri}, {La Franca}, {Maiolino}, {Matt},
  {Molendi}, {Mignoli}, {Perola}, {Severgnini}, \& {Vignali}}]{fea03}
{Fiore}, F. {et~al.} 2003, \aap, 409, 79

\bibitem[{{Flohic} {et~al.}(2006){Flohic}, {Eracleous}, {Chartas}, {Shields},
  \& {Moran}}]{fea06}
{Flohic}, H.~M.~L.~G., {Eracleous}, M., {Chartas}, G., {Shields}, J.~C., \&
  {Moran}, E.~C. 2006, \apj, 647, 140

\bibitem[{{Freedman} {et~al.}(2001){Freedman}, {Madore}, {Gibson}, {Ferrarese},
  {Kelson}, {Sakai}, {Mould}, {Kennicutt}, {Ford}, {Graham}, {Huchra},
  {Hughes}, {Illingworth}, {Macri}, \& {Stetson}}]{fea01}
{Freedman}, W.~L. {et~al.} 2001, \apj, 553, 47

\bibitem[{{Ganda} {et~al.}(2006){Ganda}, {Falc{\'o}n-Barroso}, {Peletier},
  {Cappellari}, {Emsellem}, {McDermid}, {de Zeeuw}, \& {Carollo}}]{gfea06}
{Ganda}, K., {Falc{\'o}n-Barroso}, J., {Peletier}, R.~F., {Cappellari}, M.,
  {Emsellem}, E., {McDermid}, R.~M., {de Zeeuw}, P.~T., \& {Carollo}, C.~M.
  2006, \mnras, 367, 46

\bibitem[{{Gebhardt} {et~al.}(2000){Gebhardt}, {Bender}, {Bower}, {Dressler},
  {Faber}, {Filippenko}, {Green}, {Grillmair}, {Ho}, {Kormendy}, {Lauer},
  {Magorrian}, {Pinkney}, {Richstone}, \& {Tremaine}}]{gea00}
{Gebhardt}, K. {et~al.} 2000, \apjl, 539, L13

\bibitem[{{Ghosh} {et~al.}(2007){Ghosh}, {Pogge}, {Mathur}, {Martini}, \&
  {Shields}}]{gea07}
{Ghosh}, H., {Pogge}, R.~W., {Mathur}, S., {Martini}, P., \& {Shields}, J.~C.
  2007, \apj, 656, 105

\bibitem[{{Gon{\c c}alves} {et~al.}(1999){Gon{\c c}alves}, {V{\'e}ron-Cetty},
  \& {V{\'e}ron}}]{gvv99}
{Gon{\c c}alves}, A.~C., {V{\'e}ron-Cetty}, M.-P., \& {V{\'e}ron}, P. 1999,
  \aaps, 135, 437

\bibitem[{{Graham} {et~al.}(2007){Graham}, {Driver}, {Allen}, \&
  {Liske}}]{gdea07}
{Graham}, A.~W., {Driver}, S.~P., {Allen}, P.~D., \& {Liske}, J. 2007, \mnras,
  378, 198

\bibitem[{{Graham} {et~al.}(2001){Graham}, {Erwin}, {Caon}, \&
  {Trujillo}}]{gea01}
{Graham}, A.~W., {Erwin}, P., {Caon}, N., \& {Trujillo}, I. 2001, \apjl, 563,
  L11

\bibitem[{{Greene} \& {Ho}(2004)}]{gh04}
{Greene}, J.~E., \& {Ho}, L.~C. 2004, \apj, 610, 722

\bibitem[{{Greene} \& {Ho}(2007{\natexlab{a}})}]{gh07b}
------. 2007{\natexlab{a}}, \apj, 667, 131

\bibitem[{{Greene} \& {Ho}(2007{\natexlab{b}})}]{gh07a}
------. 2007{\natexlab{b}}, \apj, 656, 84

\bibitem[{{Grimm} {et~al.}(2003){Grimm}, {Gilfanov}, \& {Sunyaev}}]{ggs03}
{Grimm}, H.-J., {Gilfanov}, M., \& {Sunyaev}, R. 2003, \mnras, 339, 793

\bibitem[{{Hasinger} {et~al.}(2005){Hasinger}, {Miyaji}, \& {Schmidt}}]{hms05}
{Hasinger}, G., {Miyaji}, T., \& {Schmidt}, M. 2005, \aap, 441, 417

\bibitem[{{Heckman}(1980)}]{h80}
{Heckman}, T.~M. 1980, \aap, 87, 152

\bibitem[{{H{\'e}raudeau} \& {Simien}(1998)}]{hs98}
{H{\'e}raudeau}, P., \& {Simien}, F. 1998, \aaps, 133, 317

\bibitem[{{Ho}(2008)}]{h08}
{Ho}, L.~C. 2008, ArXiv e-prints, 0803.2268

\bibitem[{{Ho} {et~al.}(1995){Ho}, {Filippenko}, \& {Sargent}}]{hfs95}
{Ho}, L.~C., {Filippenko}, A.~V., \& {Sargent}, W.~L. 1995, \apjs, 98, 477

\bibitem[{{Ho} {et~al.}(1997{\natexlab{a}}){Ho}, {Filippenko}, \&
  {Sargent}}]{hfs97-3}
{Ho}, L.~C., {Filippenko}, A.~V., \& {Sargent}, W.~L.~W. 1997{\natexlab{a}},
  \apjs, 112, 315

\bibitem[{{Ho} {et~al.}(1997{\natexlab{b}}){Ho}, {Filippenko}, {Sargent}, \&
  {Peng}}]{hfsp97}
{Ho}, L.~C., {Filippenko}, A.~V., {Sargent}, W.~L.~W., \& {Peng}, C.~Y.
  1997{\natexlab{b}}, \apjs, 112, 391

\bibitem[{{Holtzman} {et~al.}(1995){Holtzman}, {Hester}, {Casertano},
  {Trauger}, {Watson}, {Ballester}, {Burrows}, {Clarke}, {Crisp}, {Evans},
  {Gallagher}, {Griffiths}, {Hoessel}, {Matthews}, {Mould}, {Scowen},
  {Stapelfeldt}, \& {Westphal}}]{hea95}
{Holtzman}, J.~A. {et~al.} 1995, \pasp, 107, 156

\bibitem[{{Kennicutt} {et~al.}(2003){Kennicutt}, {Armus}, {Bendo}, {Calzetti},
  {Dale}, {Draine}, {Engelbracht}, {Gordon}, {Grauer}, {Helou}, {Hollenbach},
  {Jarrett}, {Kewley}, {Leitherer}, {Li}, {Malhotra}, {Regan}, {Rieke},
  {Rieke}, {Roussel}, {Smith}, {Thornley}, \& {Walter}}]{kea03}
{Kennicutt}, Jr., R.~C. {et~al.} 2003, \pasp, 115, 928

\bibitem[{{Kormendy} \& {Richstone}(1995)}]{kr95}
{Kormendy}, J., \& {Richstone}, D. 1995, \araa, 33, 581

\bibitem[{{Kotanyi}(1980)}]{k80}
{Kotanyi}, C.~G. 1980, \aaps, 41, 421

\bibitem[{{Laor}(2003)}]{l03}
{Laor}, A. 2003, \apj, 590, 86

\bibitem[{{Larsen}(2004)}]{l04}
{Larsen}, S.~S. 2004, \aap, 416, 537

\bibitem[{{Levenson} {et~al.}(2006){Levenson}, {Heckman}, {Krolik}, {Weaver},
  \& {{\.Z}ycki}}]{lea06}
{Levenson}, N.~A., {Heckman}, T.~M., {Krolik}, J.~H., {Weaver}, K.~A., \&
  {{\.Z}ycki}, P.~T. 2006, \apj, 648, 111

\bibitem[{{Marconi} \& {Hunt}(2003)}]{mh03}
{Marconi}, A., \& {Hunt}, L.~K. 2003, \apjl, 589, L21

\bibitem[{{Marconi} {et~al.}(2004){Marconi}, {Risaliti}, {Gilli}, {Hunt},
  {Maiolino}, \& {Salvati}}]{mea04}
{Marconi}, A., {Risaliti}, G., {Gilli}, R., {Hunt}, L.~K., {Maiolino}, R., \&
  {Salvati}, M. 2004, \mnras, 351, 169

\bibitem[{{Martini} {et~al.}(2002){Martini}, {Kelson}, {Mulchaey}, \&
  {Trager}}]{mea02}
{Martini}, P., {Kelson}, D.~D., {Mulchaey}, J.~S., \& {Trager}, S.~C. 2002,
  \apjl, 576, L109

\bibitem[{{McElroy}(1995)}]{m95}
{McElroy}, D.~B. 1995, \apjs, 100, 105

\bibitem[{{McLure} \& {Dunlop}(2001)}]{md01}
{McLure}, R.~J., \& {Dunlop}, J.~S. 2001, \mnras, 327, 199

\bibitem[{{McLure} {et~al.}(2004){McLure}, {Willott}, {Jarvis}, {Rawlings},
  {Hill}, {Mitchell}, {Dunlop}, \& {Wold}}]{mwea04}
{McLure}, R.~J., {Willott}, C.~J., {Jarvis}, M.~J., {Rawlings}, S., {Hill},
  G.~J., {Mitchell}, E., {Dunlop}, J.~S., \& {Wold}, M. 2004, \mnras, 351, 347

\bibitem[{{Merloni}(2004)}]{m04}
{Merloni}, A. 2004, \mnras, 353, 1035

\bibitem[{{Mouri} \& {Taniguchi}(1992)}]{mt92}
{Mouri}, H., \& {Taniguchi}, Y. 1992, \apj, 386, 68

\bibitem[{{Nagar} {et~al.}(2005){Nagar}, {Falcke}, \& {Wilson}}]{nfw05}
{Nagar}, N.~M., {Falcke}, H., \& {Wilson}, A.~S. 2005, \aap, 435, 521

\bibitem[{{Nagar} {et~al.}(2002){Nagar}, {Falcke}, {Wilson}, \&
  {Ulvestad}}]{nea02}
{Nagar}, N.~M., {Falcke}, H., {Wilson}, A.~S., \& {Ulvestad}, J.~S. 2002, \aap,
  392, 53

\bibitem[{{Neff} \& {Ulvestad}(2000)}]{nu00}
{Neff}, S.~G., \& {Ulvestad}, J.~S. 2000, \aj, 120, 670

\bibitem[{{Nicastro}(2000)}]{n00}
{Nicastro}, F. 2000, \apjl, 530, L65

\bibitem[{{Nicastro} {et~al.}(2003){Nicastro}, {Martocchia}, \& {Matt}}]{nmm03}
{Nicastro}, F., {Martocchia}, A., \& {Matt}, G. 2003, \apjl, 589, L13

\bibitem[{{Park} {et~al.}(2006){Park}, {Kashyap}, {Siemiginowska}, {van Dyk},
  {Zezas}, {Heinke}, \& {Wargelin}}]{pea06}
{Park}, T., {Kashyap}, V.~L., {Siemiginowska}, A., {van Dyk}, D.~A., {Zezas},
  A., {Heinke}, C., \& {Wargelin}, B.~J. 2006, \apj, 652, 610

\bibitem[{{Pence} {et~al.}(2001){Pence}, {Snowden}, {Mukai}, \&
  {Kuntz}}]{pea01}
{Pence}, W.~D., {Snowden}, S.~L., {Mukai}, K., \& {Kuntz}, K.~D. 2001, \apj,
  561, 189

\bibitem[{{Peterson} {et~al.}(2005){Peterson}, {Bentz}, {Desroches},
  {Filippenko}, {Ho}, {Kaspi}, {Laor}, {Maoz}, {Moran}, {Pogge}, \&
  {Quillen}}]{pea05}
{Peterson}, B.~M. {et~al.} 2005, \apj, 632, 799

\bibitem[{{Pogge}(1988)}]{p88}
{Pogge}, R.~W. 1988, \apj, 332, 702

\bibitem[{{Pounds} {et~al.}(2004){Pounds}, {Reeves}, {Page}, \&
  {O'Brien}}]{pea04}
{Pounds}, K.~A., {Reeves}, J.~N., {Page}, K.~L., \& {O'Brien}, P.~T. 2004,
  \apj, 605, 670

\bibitem[{{Prestwich} {et~al.}(2003){Prestwich}, {Irwin}, {Kilgard}, {Krauss},
  {Zezas}, {Primini}, {Kaaret}, \& {Boroson}}]{pea03}
{Prestwich}, A.~H., {Irwin}, J.~A., {Kilgard}, R.~E., {Krauss}, M.~I., {Zezas},
  A., {Primini}, F., {Kaaret}, P., \& {Boroson}, B. 2003, \apj, 595, 719

\bibitem[{{Ranalli} {et~al.}(2003){Ranalli}, {Comastri}, \& {Setti}}]{rcs03}
{Ranalli}, P., {Comastri}, A., \& {Setti}, G. 2003, \aap, 399, 39

\bibitem[{{Randall} {et~al.}(2006){Randall}, {Sarazin}, \& {Irwin}}]{rsi06}
{Randall}, S.~W., {Sarazin}, C.~L., \& {Irwin}, J.~A. 2006, \apj, 636, 200

\bibitem[{{Satyapal} {et~al.}(2004){Satyapal}, {Sambruna}, \& {Dudik}}]{ssd04}
{Satyapal}, S., {Sambruna}, R.~M., \& {Dudik}, R.~P. 2004, \aap, 414, 825

\bibitem[{{Satyapal} {et~al.}(2008){Satyapal}, {Vega}, {Dudik}, {Abel}, \&
  {Heckman}}]{sea08}
{Satyapal}, S., {Vega}, D., {Dudik}, R.~P., {Abel}, N.~P., \& {Heckman}, T.
  2008, ArXiv e-prints, 0801.2759

\bibitem[{{Satyapal} {et~al.}(2007){Satyapal}, {Vega}, {Heckman}, {O'Halloran},
  \& {Dudik}}]{svea07}
{Satyapal}, S., {Vega}, D., {Heckman}, T., {O'Halloran}, B., \& {Dudik}, R.
  2007, \apjl, 663, L9

\bibitem[{{Seth} {et~al.}(2008){Seth}, {Agueros}, {Lee}, \&
  {Basu-Zych}}]{salb08}
{Seth}, A., {Agueros}, M., {Lee}, D., \& {Basu-Zych}, A. 2008, \apj, in press

\bibitem[{{Shankar} {et~al.}(2004){Shankar}, {Salucci}, {Granato}, {De Zotti},
  \& {Danese}}]{sea04}
{Shankar}, F., {Salucci}, P., {Granato}, G.~L., {De Zotti}, G., \& {Danese}, L.
  2004, \mnras, 354, 1020

\bibitem[{{Shields} {et~al.}(2008){Shields}, {Walcher}, {Boeker}, {Ho}, {Rix},
  \& {van der Marel}}]{swea08}
{Shields}, J.~C., {Walcher}, C.~J., {Boeker}, T., {Ho}, L.~C., {Rix}, H.-W., \&
  {van der Marel}, R.~P. 2008, \apj, in press

\bibitem[{{Skrutskie} {et~al.}(2006){Skrutskie}, {Cutri}, {Stiening},
  {Weinberg}, {Schneider}, {Carpenter}, {Beichman}, {Capps}, {Chester},
  {Elias}, {Huchra}, {Liebert}, {Lonsdale}, {Monet}, {Price}, {Seitzer},
  {Jarrett}, {Kirkpatrick}, {Gizis}, {Howard}, {Evans}, {Fowler}, {Fullmer},
  {Hurt}, {Light}, {Kopan}, {Marsh}, {McCallon}, {Tam}, {Van Dyk}, \&
  {Wheelock}}]{2MASS}
{Skrutskie}, M.~F. {et~al.} 2006, \aj, 131, 1163

\bibitem[{{Snellen} {et~al.}(2003){Snellen}, {Lehnert}, {Bremer}, \&
  {Schilizzi}}]{sea03}
{Snellen}, I.~A.~G., {Lehnert}, M.~D., {Bremer}, M.~N., \& {Schilizzi}, R.~T.
  2003, \mnras, 342, 889

\bibitem[{{Soria} {et~al.}(2007){Soria}, {Baldi}, {Risaliti}, {Fabbiano},
  {King}, {La Parola}, \& {Zezas}}]{sea07}
{Soria}, R., {Baldi}, A., {Risaliti}, G., {Fabbiano}, G., {King}, A., {La
  Parola}, V., \& {Zezas}, A. 2007, \mnras, 379, 1313

\bibitem[{{Soria} {et~al.}(2006{\natexlab{a}}){Soria}, {Fabbiano}, {Graham},
  {Baldi}, {Elvis}, {Jerjen}, {Pellegrini}, \& {Siemiginowska}}]{sfea06}
{Soria}, R., {Fabbiano}, G., {Graham}, A.~W., {Baldi}, A., {Elvis}, M.,
  {Jerjen}, H., {Pellegrini}, S., \& {Siemiginowska}, A. 2006{\natexlab{a}},
  \apj, 640, 126

\bibitem[{{Soria} {et~al.}(2006{\natexlab{b}}){Soria}, {Graham}, {Fabbiano},
  {Baldi}, {Elvis}, {Jerjen}, {Pellegrini}, \& {Siemiginowska}}]{sgea06}
{Soria}, R., {Graham}, A.~W., {Fabbiano}, G., {Baldi}, A., {Elvis}, M.,
  {Jerjen}, H., {Pellegrini}, S., \& {Siemiginowska}, A. 2006{\natexlab{b}},
  \apj, 640, 143

\bibitem[{{Stern} {et~al.}(2005){Stern}, {Eisenhardt}, {Gorjian}, {Kochanek},
  {Caldwell}, {Eisenstein}, {Brodwin}, {Brown}, {Cool}, {Dey}, {Green},
  {Jannuzi}, {Murray}, {Pahre}, \& {Willner}}]{skea05}
{Stern}, D. {et~al.} 2005, \apj, 631, 163

\bibitem[{{Stetson} {et~al.}(1998)}]{ssea98}
{Stetson}, P.~B., {et~al.} 1998, \apj, 508, 491

\bibitem[{{Stobbart} {et~al.}(2006){Stobbart}, {Roberts}, \& {Wilms}}]{srw06}
{Stobbart}, A.-M., {Roberts}, T.~P., \& {Wilms}, J. 2006, \mnras, 368, 397

\bibitem[{{Terashima} \& {Wilson}(2003)}]{tw03}
{Terashima}, Y., \& {Wilson}, A.~S. 2003, \apj, 583, 145

\bibitem[{{Tully}(1988)}]{t88}
{Tully}, R.~B. 1988, {Nearby galaxies catalog} (Cambridge and New York,
  Cambridge University Press, 1988, 221 p.)

\bibitem[{{Tzanavaris} \& {Georgantopoulos}(2007)}]{tg07}
{Tzanavaris}, P., \& {Georgantopoulos}, I. 2007, \aap, 468, 129

\bibitem[{{Ulvestad} \& {Ho}(2002)}]{uh02}
{Ulvestad}, J.~S., \& {Ho}, L.~C. 2002, \apj, 581, 925

\bibitem[{{Veilleux} \& {Osterbrock}(1987)}]{vo87}
{Veilleux}, S., \& {Osterbrock}, D.~E. 1987, \apjs, 63, 295

\bibitem[{{Walcher} {et~al.}(2005){Walcher}, {van der Marel}, {McLaughlin},
  {Rix}, {B{\"o}ker}, {H{\"a}ring}, {Ho}, {Sarzi}, \& {Shields}}]{wea05}
{Walcher}, C.~J. {et~al.} 2005, \apj, 618, 237

\bibitem[{{Williams} {et~al.}(2004){Williams}, {Mathur}, \& {Pogge}}]{wmp04}
{Williams}, R.~J., {Mathur}, S., \& {Pogge}, R.~W. 2004, \apj, 610, 737

\bibitem[{{Willis} {et~al.}(1976){Willis}, {Oosterbaan}, \& {de
  Ruiter}}]{wod76}
{Willis}, A.~G., {Oosterbaan}, C.~E., \& {de Ruiter}, H.~R. 1976, \aaps, 25,
  453

\bibitem[{{Young} {et~al.}(2007){Young}, {Nowak}, {Markoff}, {Marshall}, \&
  {Canizares}}]{yea07}
{Young}, A.~J., {Nowak}, M.~A., {Markoff}, S., {Marshall}, H.~L., \&
  {Canizares}, C.~R. 2007, \apj, 669, 830

\end{thebibliography}
\end{document}